\newif\iftwocolumn
\newif\ifonecolumn
\newif\iflncs
\newif\ifccs
\newif\ifanonymous
\newif\iftoday
\todaytrue
\lncstrue

\ifccs
  \twocolumntrue
\fi
\iflncs
  \onecolumntrue
\fi

\iflncs
\documentclass[11pt]{llncs}
\fi
\ifccs
\documentclass[]{acmart}
\input{ccs-preamble}
\fi

\makeatletter
\renewcommand\subsubsection{\@startsection{subsubsection}{3}{\z@}%
                       {-18\p@ \@plus -4\p@ \@minus -4\p@}%
                       {0.5em \@plus 0.22em \@minus 0.1em}%
                       {\normalfont\normalsize\bfseries\boldmath}}
\makeatother
\usepackage{preamble}
\usepackage{graphicx}
\usepackage{tablefootnote}
\usepackage{pythonhighlight}
\usepackage[capitalise,nameinlink,noabbrev,compress]{cleveref}

\usepackage{listings, listings-rust}
\usepackage[normalem]{ulem}
\usepackage{slantsc}
\DeclareFontShape{OT1}{cmr}{m}{scit}{<->ssub*cmr/m/sc}{}

\usepackage{geometry}
\begin{document}

\title{
  Formalization and security analysis of \\the Bridgeless protocol \thanks{
    This work was funded by \btm (\url{https://bridgeless.com/}).
    The following sources were used for the protocol's description:
        (1) The \btm docs: \url{https://docs.devnet.bridgeless.com/docs/specs/intro}
        (2) The \emph{tss-svc} repository: \url{https://github.com/hyle-team/tss-svc/tree/2373e34705592388cf262d500c4cf07798765d5a}
        (3) The \emph{bridge-evm-contracts} repository: \url{https://github.com/dl-tokene/bridge-evm-contracts/tree/d67aee0e107971de2ca8ecfa07c4c3f70ebc5e96}
    }
}
\ifanonymous{\iflncs
\fi}
\else
\author{
  Orestis Alpos\inst{1}\and
  Oleg Fomenko\inst{2}\and
  Dimitris Karakostas\inst{1,3}\and\\
  Oleksandr Kurbatov\inst{2}\and
  Andrey Sabelnikov\inst{4}
}
\iflncs
\institute{
  Common Prefix
  \and
  Distributed Lab
  \and
  University of Edinburgh
  \and
  Zano Project
}
\else
\affiliation{
\institution{
}
}
\fi
\fi

\iflncs
\maketitle
\fi

\iftoday
\noindent
\makebox[\linewidth]{Last update: \today}
\fi

\begin{abstract}

This paper formalizes the proves the security of the \btm protocol,
a protocol able to bridge tokens between various chains. The \btm protocol is run by a set of validators, responsible for verifying deposit transactions on the source chain and generating the corresponding withdrawals on the target chain. The protocol is designed to be chain-agnostic and the validators interact with each supported chain via a \emph{chain client}.
It currently supports EVM-compatible chains, the Zano, and the Bitcoin chains.
The paper formalizes all involved subprotocols and describes the conditions under which the protocol maintains safety and liveness.

\end{abstract}

\ifccs
\input{ccs-keywords}
\maketitle
\fi

\section{Introduction}

This paper describes \btm, a protocol for bridging tokens between different
blockchains. \btm supports both programmable ledgers, like Ethereum, and
ledgers without smart contract capabilities, like Bitcoin.

At a high level, token bridging occurs via a \emph{bridging request} (or simply
\emph{request}). Intuitively, a request is the sequence of actions that starts
with a deposit on the source chain and ends with the corresponding withdrawal
on the target chain. The \emph{deposit} is the action of making an amount of
tokens unavailable on the source chain, following a specific mechanism defined
by the bridge protocol, such as locking them to a specific address or burning
them. Following, the \emph{withdrawal} makes the corresponding amount of tokens
available on the target chain, following a specific mechanism defined by the
bridge protocol, like transferring them from a specific address or minting
them.

The rest of this paper is organized as follows. First, \cref{sec:overview}
outlines the lifecycle of a bridging request and the bridge's main phases and
components. Next, \cref{sec:model} details the model, under which we will
describe and analyze the protocol, and useful primitives from the literature
that will be used in the construction. \cref{sec:bridge-protocol} offers the
main contribution of this work, which is the precise description of \btm.
Following, \cref{sec:analysis} provides the formal analysis of the protocol,
showing that \btm guarantees a set of useful properties, like bridge safety and
liveness. Finally, Sections \ref{sec:deposit}{-}\ref{sec:chain-clients} discuss
implementation details regarding the deposit, withdrawal, and collecting and
publishing the necessary objects from the source and target ledgers.

\section{Overview}\label{sec:overview}

\begin{figure}
    \center
    \includegraphics[width=0.9\columnwidth]{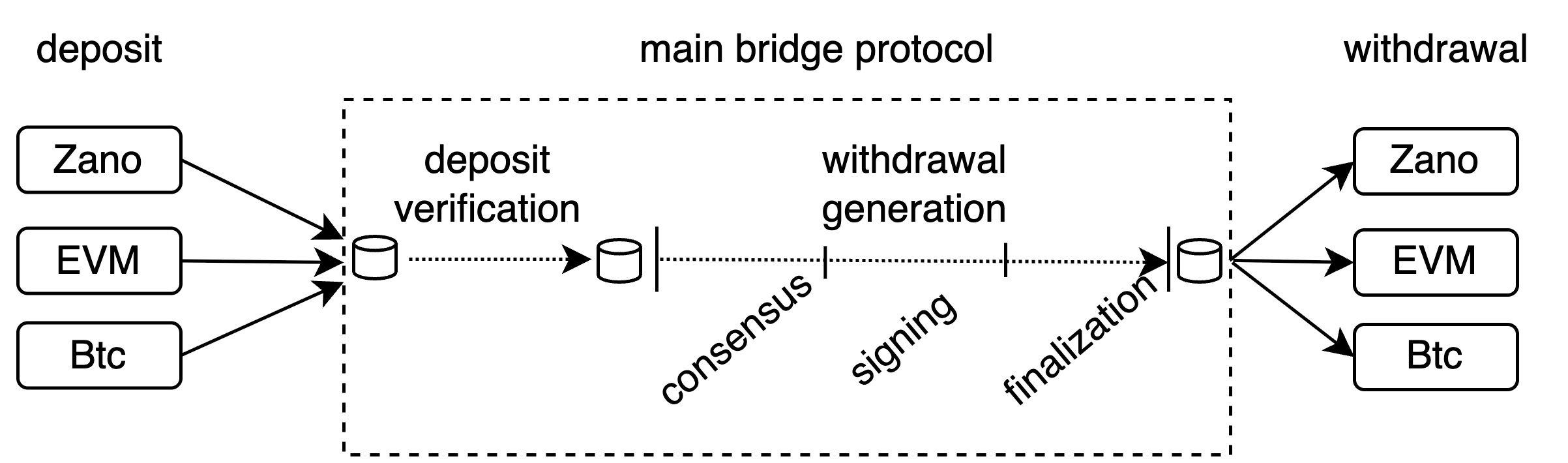}
    \caption{The flow of a bridging request.}
    \label{fig:protocol-overview}
\end{figure}

In this section we give an overview of the lifecycle of a bridging request. The
flow is depicted in \Cref{fig:protocol-overview}.

\dotparagraph{Deposit}
A bridging request is initiated by the client performing a \emph{deposit} on the source chain.
After the deposit has been finalized, according to the rules of the source chain, the client is responsible for forwarding the required information to the \btm validators. To that end, the client creates a \emph{deposit identifier} and sends it to the validators. We show the details for each supported chain in \Cref{sec:deposit}.

\dotparagraph{Main bridge protocol}
When a validator receives a bridging request, it stores it locally and processes it using the \btm protocol.
The main bridge protocol is executed by the \btm validators. It consists of two subprotocols.
\begin{enumerate}
    \item \emph{Deposit verification}: The goal of the subprotocol is to verify that the deposit indicated by the deposit identifier has indeed happened at the source chain. Each validator \emph{independently} verifies the deposit and, if successful, \emph{locally} marks the request as verified and stores the deposit data. We show the details in \Cref{sec:bridge-deposit-verification}.
    \item \emph{Withdrawal generation}: The goal of the subprotocol is to agree on the request to be processed and generate a withdrawal transaction and a signature for the target chain. This is an \emph{interactive} protocol run by the validators, using one validator as the \emph{proposer}. It consists of three sub-phases, named \emph{consensus}, \emph{signature generation}, and \emph{finalization}, each with a fixed duration, run one after the other. We show the details in \Cref{sec:bridge-withdrawal-generation}.
\end{enumerate}

\dotparagraph{Withdrawal}
Finally, the withdrawal transaction, which has been generated by the withdrawal-generation protocol, is submitted to the target chain.
Here the protocol differs depending on the target chain.
If the target chain is Zano or Bitcoin, the validators submit the withdrawal transaction.
If the target is an EVM chain, the client is responsible for submitting the generated withdrawal transaction. We show the details in \Cref{sec:withdrawal}.

\medskip
Each request is associated with a \var{status}, which a validator updates during the execution of the protocol as follows:
\begin{itemize}
    \item \const{invalid}: The validator has received the request from a client.
    \item \const{pending}: The validator has verified the deposit transaction on the source chain.
    \item \const{processing}: The validator has delivered a \msgsignstartEMPTY message  for that request (defined in \Cref{sec:consensus-phase}).
    \item \const{processed}: The validator has obtained a valid signature for the withdrawal transaction that corresponds to the request.
    \item \const{finalized}: The validator has observed the withdrawal transaction finalized on the target chain.
\end{itemize}

\begin{figure}
    \center
    \includegraphics[width=0.9\columnwidth]{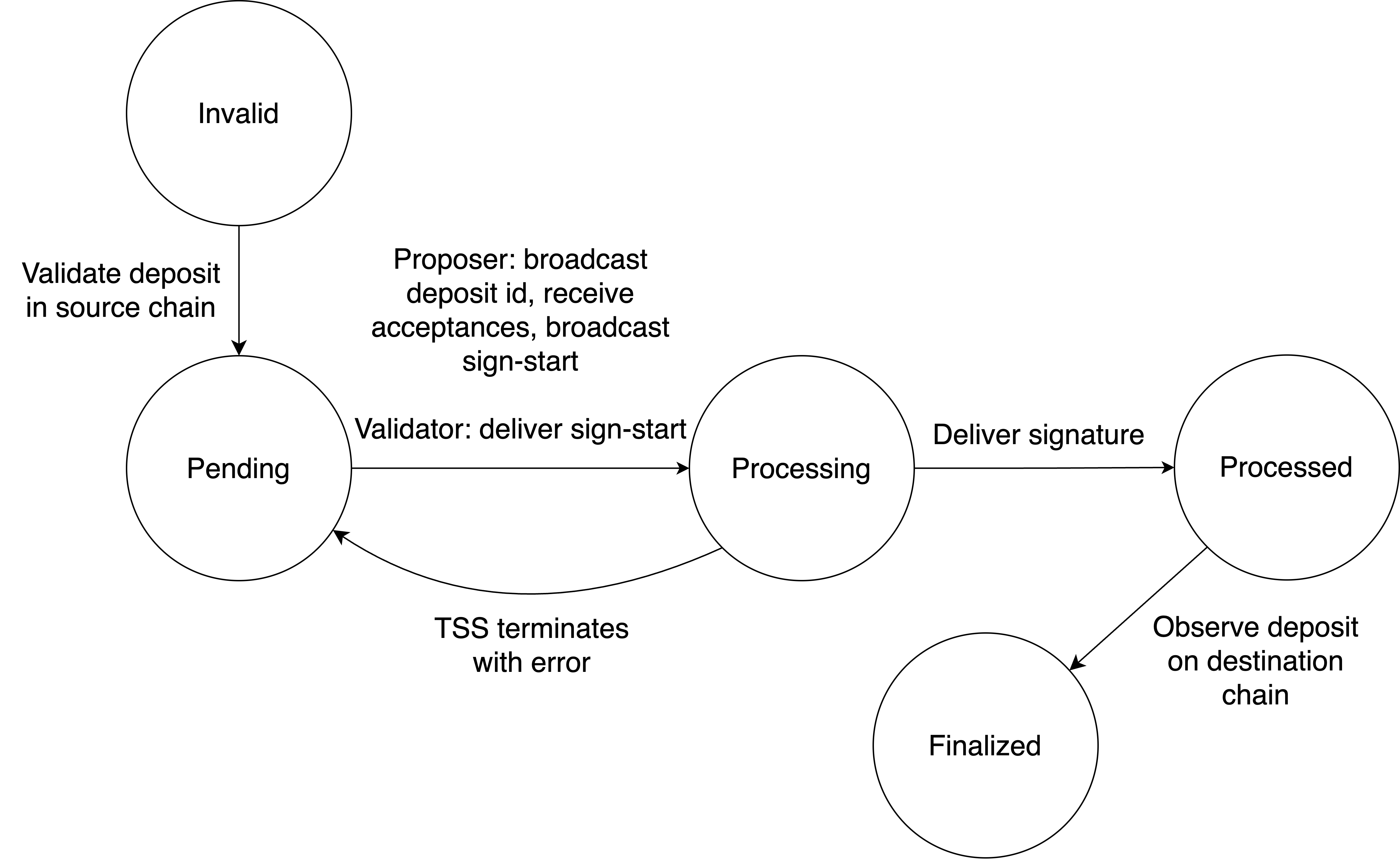}
    \caption{The state machine of a bridgeless validator during the protocol execution.}
    \label{fig:state-machine}
\end{figure}

The \var{status} hence defines a \emph{state machine} executed by each validator, which we also depict in \Cref{fig:state-machine}. The security analysis of the protocol (presented in \Cref{sec:analysis}) is built on the \var{status} of a request. We show that if an honest client initiates a bridging request, then honest validators will eventually see the request as \const{finalized} (liveness of the bridge), and that if the withdrawal appears in the target chain then honest validators have previously marked the request as \const{pending}, hence the deposit has been finalized in the source chain (safety of the bridge).

\section{Model and Preliminaries}\label{sec:model}

\dotparagraph{Bridge}
In this work we consider a \emph{bridge} as an interoperability protocol between
a \emph{source chain} and a \emph{target chain}, with the purpose to move
\emph{assets} from the source to the target chain. Without loss of generality we
assume it is unidirectional. A bridge satisfies the properties of \emph{bridge
liveness} and \emph{bridge safety} (Definitions \ref{def:bridge-liveness} and
\ref{def:bridge-safety}).

\begin{definition}[Bridge liveness]\label{def:bridge-liveness}
    A bridge protocol satisfies \emph{bridge liveness} with parameter $r$ if every bridging request made by a honest party is completed after at most $r$ rounds.
    In other words, for every honest deposit in the source chain, the corresponding withdrawal in the target chain is eventually executed.
\end{definition}

\begin{definition}[Bridge safety]\label{def:bridge-safety}
    A bridge protocol satisfies \emph{bridge safety} if, for every withdrawal in the target chain, there exists a corresponding deposit in the source chain.
\end{definition}

\dotparagraph{Parties}
The following parties participate in the protocol:
\begin{enumerate}
    \item \emph{Validators}: A set of $n$ validators, $v_1, \ldots, v_n$, out of which $t$ are malicious, are responsible for running the bridge protocol, each having equal weight in the voting and signing schemes.
    \item \emph{Clients}: They initiate bridging requests and also act as \emph{relayers}, responsible for relaying certain information from the source chain to the validators and from the validators to the target chain.
    \item \emph{Ethereum RPC provider}: A public provider (e.g., Infura) used by the validators to query the state of the Ethereum chain.
\end{enumerate}

\dotparagraph{Corruption model}
Malicious validators can arbitrarily deviate from the protocol. We assume that, out of the $n$ validators, at most $t = \lfloor n/3 \rfloor$ are malicious.

\dotparagraph{Network assumptions}
The network is assumed to be \emph{synchronous}. Certain phases of the protocol have a fixed duration, and all messages by honest parties are assumed to arrive within that duration.

\dotparagraph{Pseudocode notation} The notation `\textbf{require} \var{condition}' causes a function to terminate immediately and return \error, if \var{condition} is false.

Our work makes use of the following cryptographic primitives from the literature.

\begin{definition}[Reliable Broadcast (RB)]\label{def:reliable-broadcast}
    A reliable broadcast primitive has the following interface:
    \begin{itemize}
        \item function \var{broadcast}(\var{message})
        \item function \var{deliver}(\var{message}, \var{sender})
    \end{itemize}
    It satisfies the following properties:
    \begin{itemize}
        \item \emph{Termination}: Every honest validator delivers some value.
        \item \emph{Validity}: If an honest validator broadcasts a message $m$, then every honest validator eventually delivers $m$.
        \item \emph{Integrity}: If an honest validator delivers $m$ with sender $p$, then $p$ has previously broadcast $m$.
        \item \emph{Agreement}: If an honest validator delivers a message $m$, then all honest validators deliver $m$.
    \end{itemize}
\end{definition}

Each instance of a reliable broadcast protocol, denoted by $\var{RB}[\var{id}]$ and parametrized by an identifier $\var{id}$, is run among a pre-determined set of participants.

\begin{definition}[(Binance) Threshold Signature Scheme (TSS)]\label{def:binance-tss}
    The Threshold Signature Scheme
    is run by a set of $m$ parties and satisfies the following properties:
    \begin{itemize}
        \item \emph{Termination:} If at least one signer is honest, then the primitive terminates.
        \item \emph{Completeness:} If all parties are honest, then the signature produced by them is valid, i.e., it verifies successfully with overwhelming probability.
        \item \emph{Agreement:} All honest parties receive the same TSS output.
        \item \emph{Unforgeability:} If a valid signature is output, then all $m$ parties have signed the message.
    \end{itemize}
\end{definition}

Optionally, the TSS primitive may also offer \emph{accountability}. In this
case, if the signature generation process is invalid, that is if the returned
signature does not verify successfully, then the honest signers can identify at
least one signer as misbehaving. However, as we mentioned above, accountability
is outside the scope of this work.

\section{Bridgeless Protocol}\label{sec:bridge-protocol}
In this section we describe the two subprotocols of \btm which are run by the validators.

Note that a bridging request is initiated by a \emph{deposit} transaction, submitted by a client on the source chain. In this section we assume an abstract deposit logic, and in \Cref{sec:deposit} we show the exact implementation. Moreover, a bridging request is terminated by a \emph{withdrawal} transaction, performed by the validators on the target chain. In this section we also treat the withdrawal format as abstract, and show in \Cref{sec:withdrawal} the implementation details.

Note also that the validators of \btm are required to interact with the supported chains in order to fetch events from the source chain (in the deposit-verification subprotocol) and generate transactions for the target chain (in the signing subprotocol). The module of a validator that is responsible for this is called a \emph{\ChainClient}.
In this section we treat the \ChainClient as a black box which exposes the functions \op{getDepositData()} and \op{getHashOfWithdrawal()}. The inner workings of the \ChainClient are described in \Cref{sec:chain-clients}.

In \Cref{alg:data-types} we define the data types and the state used by the validators throughout the protocol.
A validator maintains all bridging requests it is aware of in a variable \var{requests}, which is a map from a \var{depositId} to some \type{RequestData}.
Observe that \type{RequestData} consists of two parts, the \type{DepositData} and the \type{WithdrawalData}.
The first will contain information related to the deposit of bridging request, read from the deposit transaction on the source chain, while the second will be filled with information required for submitting the withdrawal transaction on the target chain.

\begin{algorithm}[ht!]
\caption{Data types used by \btm validators.}
\label{alg:data-types}
\begin{algorithmic}[1]
    \State{type \type{DepositIdentifier}: \{\var{txHash}, \var{txNonce}, \var{chainID}\}} \Comment{Unique identifier for each deposit}
    \State{type \type{DepositData}: \{\var{tokenAddr}, \var{amount}, \var{sourceAddr}, \var{targetAddr}, \var{targetChain}\}} \Comment{Read in deposit tx}
    \State{type \type{WithdrawalData}: \{\var{signHash}, \var{signers}, \var{signature}\}} \Comment{Constructed by validators for withdrawal tx}
    \State{type \type{Status}: \{\const{invalid}, \const{pending}, \const{processing}, \const{processed}, \const{finalized}\}}
    \State{type \type{RequestData}: \{\var{status}: \type{Status}, \var{depositData}: \type{DepositData}, \var{withdrawalData}: \type{WithdrawalData}\}}
    \State{var \var{requests}: $\type{DepositIdentifier} \rightarrow \type{RequestData}$}
    \State{var \var{bridgeValidators}} \Comment{The validators of \btm}
    \State{const \const{threshold}} \Comment{used in the consensus phase, instantiated with $\const{threshold} = 2t$}
    \Let{\var{lockedBitcoinInputs}}{\emptyset} \label{line:consensus-init-locked} \Comment{already used bitcoin UTXOs, see \Cref{alg:finalization-phase} and \Cref{alg:btc-chain-client}}
\end{algorithmic}
\end{algorithm}

\subsection{Deposit verification} \label{sec:bridge-deposit-verification}
The deposit verification subprotocol is run by the \btm validators independently and locally, for each bridging request initiated by a client. It is initiated when the validator receives a \msgsubmitwithrawal{} message from the client or another validator (in case the client sends the request only to a subset of validators).

\begin{algorithm}[ht!]
\caption{Validator code, deposit-verification protocol.}
\label{alg:tss}
\begin{algorithmic}[1]
    \On{$\msgsubmitwithrawal{\var{depositIdentifier}} \leftarrow \text{party } p \in \var{bridgeValidators} \cup \{ c \}$}
        \State{\op{submitWithdrawal}(\var{depositIdentifier})}
    \EndOn
    \On{$\msgcheckwithrawal{\var{depositIdentifier}} \leftarrow \text{client } c$}
        \State{\op{checkWithdrawal}(\var{depositIdentifier}, $c$)}
    \EndOn
    \Statex

    \Function{\op{submitWithdrawal(\var{depositIdentifier})}}{}
        \If{$\var{requests}[\var{depositIdentifier}] \neq \bot$}
            \State{\textbf{return}} \Comment{ignore duplicate requests}
        \EndIf
        \Guard{\var{depositIdentifier.chainID} \text{ is a supported chain}}
        \Let{\sourceClient}{\text{the } \ChainClient \text{ responsible for }\var{depositIdentifier.chainID}} \label{line:evm-retrieve-source-client}
        \Let{\var{depositData}}{\sourceClient.\op{getDepositData}(\var{depositIdentifier}}) \Comment{See Sec.~\ref{sec:chain-clients}}\label{line:evm-call-get-deposit}
        \Guard{\var{depositData} \neq \error}
        \Guard{\var{depositData.targetChainID} \text{ is a supported chain}}
        \Let{\targetClient}{\text{the \ChainClient for } \var{depositData.targetChainID} } \label{line:evm-retrieve-target-client}
        \Guard{\targetClient.\op{addressValid}(\var{depositData}.\var{targetAddr})}\label{line:evm-target-check-adress}
        \Guard{\targetClient.\op{amountValid}(\var{depositData.amount})} \label{line:evm-target-check-amount}
        \Let{\var{status}}{\const{pending}}
        \Let{\var{withdrawalData}}{\bot}
        \Let{\var{request}}{\text{\{\var{status}, \var{depositData}, \var{withdrawalData}\}}}\label{line:evm-request-data}
        \Let{\var{requests}[\var{depositIdentifier}]}{\var{request}} \label{line:evm-store-request}
        \For{\var{v} \textbf{in} \var{bridgeValidators}}\label{line:evm-echo-withdrawal-begin}
            \State{$\msgsubmitwithrawal{\var{depositIdentifier}} \rightarrow v$}
        \EndFor\label{line:evm-echo-withdrawal-end}
    \EndFunction
    \Statex
    \Function{\op{checkWithdrawal(\var{depositIdentifier}, $c$)}}{}\label{line:check-withdrawal}
        \Let{\var{request}}{\var{requests}[\var{depositIdentifier}]}
        \Guard{\var{request}.\var{status} == \const{processed}}
        \State{$\msgcheckwithrawalresponse{\var{request}.\var{depositData}, \var{request}.\var{withdrawalData}.\var{signature}} \rightarrow c$}
    \EndFunction
    \Statex
\end{algorithmic}
\end{algorithm}

When \msgsubmitwithrawal{} is received, the validators run \op{submitWithdrawal()}. This function first makes some basic validation, such as ignoring duplicate deposits and verifying that the source chain is supported.
After that, it retrieves the \ChainClient for the source chain (\cref{line:evm-retrieve-source-client})
(we explain the inner workings of the chain client in later sections).
The validator then uses the \sourceClient to retrieve the actual \var{depositData} (\cref{line:evm-call-get-deposit}).
If the \sourceClient indeed finds the deposit, the validator also retrieves the \ChainClient for the target chain (\cref{line:evm-retrieve-target-client}), which is used to perform some basic validation on the target address and amount of the bridging request (lines \ref{line:evm-target-check-adress}--\ref{line:evm-target-check-amount}, the function implementation is omitted).
If all checks pass, the request is marked as \const{pending} and stored locally by the validator (\cref{line:evm-store-request}).
Finally, the validator propagates the \op{submitWithdrawal()} call to all other validators (lines \ref{line:evm-echo-withdrawal-begin}--\ref{line:evm-echo-withdrawal-end}), to account for cases where the client did not send it to all validators.

\sloppy{
Finally, the validators listen to a second type of messages from clients, the \msgcheckwithrawalEMPTY{}{}, which triggers the execution of \op{checkWithdrawal()} (\cref{line:check-withdrawal}). This function allows clients to obtain the \var{withdrawalData} for a given \var{\var{depositId}}. If the corresponding request has been processed by the \btm validators, this will contain the requited signature for the withdrawal transaction.
}

\subsection{Withdrawal generation} \label{sec:bridge-withdrawal-generation}


\begin{algorithm}[ht!]
\caption{Withdrawal-generation protocol. Code for validator \var{v} and session \var{sid}.}
\label{alg:withdrawal-generation}
\begin{algorithmic}[1]
    \Function{\op{runSession}(\var{sid})}{}
        \Let{\var{proposer}}{\op{determineProposer(\var{sid})}}

        \Statex\CommentLine{Consensus phase}
        \Let{\var{consensusResult}}{\bot} \label{line:consensus-phase-start}
        \State{\textbf{start timer} \var{consensusEnd} \textbf{with duration} \const{consensusBoundary}}
        \If{\var{v} == \var{proposer}}
            \Let{\var{consensusResult}}{\op{propose}()} \Comment{See \Cref{sec:consensus-phase} and \Cref{alg:consensus}}
        \Else
            \Let{\var{consensusResult}}{\op{accept}()} \Comment{See \Cref{sec:consensus-phase} and \Cref{alg:consensus-validator}}
        \EndIf
        \On{\var{consensusEnd}}
            \Guard{\var{consensusResult} \neq \bot \text{ and } \var{consensusResult} \neq \const{error}}
            \Let{(\var{depositId}, \var{signers}, \var{signHash})}{\var{consensusResult}} \label{line:obtain-consensus-result}
            \Let{\var{requests}[\var{depositId}].\var{withdrawalData}.\var{signHash}}{\var{signHash}}
            \Let{\var{requests}[\var{depositId}].\var{withdrawalData}.\var{signers}}{\var{signers}}
            \Let{\var{requests}[\var{depositId}].\var{status}}{\const{processing}} \label{line:mark-processing}
        \EndOn\label{line:consensus-phase-end}

        \Statex\CommentLine{Signing phase}
        \Let{\var{signResult}}{\bot} \label{line:signing-phase-start}
        \State{\textbf{start timer} \var{signingEnd} \textbf{with duration} \const{signBoundary}}
        \Let{\var{signResult}}{\op{sign}(\var{signHash}, \var{signers})} \Comment{See \Cref{sec:signing-phase} and \Cref{alg:signing-phase}}
        \On{\var{signingEnd}} \label{line:signing-phase-timeout-end}
            \If{$\var{signResult} \neq \bot$}\label{line:check-error-sig}
                \Let{\var{signature}}{\var{signResult}}
                \Let{\var{requests}[\var{depositId}].\var{withdrawalData}.\var{signature}}{\var{signature}}
                \Let{\var{requests}[\var{depositId}].\var{status}}{\const{processed}} \label{line:mark-processed}
            \Else
                \Let{\var{requests}[\var{depositId}].\var{status}}{\const{pending}}
                \State{\textbf{return}} \Comment{If processing fails, then don't go to finalization.}
            \EndIf
        \EndOn\label{line:signing-phase-end}

        \Statex\CommentLine{Finalization phase}
        \Let{\var{finalizationResult}}{\bot} \label{line:finalization-phase-start}
        \State{\textbf{start timer} \var{finalizationEnd} \textbf{with duration} \const{finalizationBoundary}}
            \Let{\var{finalizationResult}}{\op{finalize}(\var{depositId}, \var{requests}[\var{depositId}].\var{depositData}, \var{signature})} \Comment{\Cref{alg:finalization-phase}}
            \State{\textbf{wait for} \var{finalizationEnd}}
            \If{$\var{finalizationResult} \neq \bot \text{ and } \var{finalizationResult} \neq \const{error}$}
                \Let{\var{requests}[\var{depositId}].\var{status}}{\const{finalized}}
            \Else
                \Let{\var{requests}[\var{depositId}].\var{status}}{\const{pending}}
            \EndIf\label{line:finalization-phase-end}
    \EndFunction

    \Statex

\end{algorithmic}
\end{algorithm}

The goal of an instance of the \emph{withdrawal generation} protocol is to agree on one request (indicated by its \var{depositId}), and generate and sign the corresponding withdrawal transaction.
The protocol proceeds in consecutive, non-overlapping \emph{sessions}, each handling one bridging request. Each session gets assigned a unique and incremental \var{sid}, has a fixed duration \const{sessionDuration}, has a well-defined validator as the \emph{proposer}. The proposer is inferred deterministically from \emph{sid}, hence the proposer for each session is known to all parties.

Each session consists of three phases, \emph{consensus} (\Cref{sec:consensus-phase}), \emph{signature generation} (\Cref{sec:signing-phase}), and \emph{finalization} (\Cref{sec:finalization-phase}), each with a fixed duration.
The timeouts used by the protocol and their current values are:
\begin{description}
    \item \const{consensusBoundary}: Duration of the consensus phase (10 sec).
    \item \const{BoundarySign}: Duration of the signature-generation phase (10 sec).
    \item \const{BoundaryFinalize}: Duration of the finalization phase (10 sec).
    \item \const{acceptanceBoundary}: Duration of the acceptance sub-phase, used by the proposer in the consensus phase (5 sec).
\end{description}

\Cref{alg:withdrawal-generation} shows the high-level pseudocode of the withdrawal generation protocol. The consensus phase (lines \ref{line:consensus-phase-start} -- \ref{line:consensus-phase-end}) returns, if successful, the \var{depositId} of the request that will be processed in the current session, as well as the \var{signHash} that needs to be signed for the withdrawal transaction, and the \var{signers} that will be responsible for signing it.
The signature-generation phase (lines \ref{line:signing-phase-start} -- \ref{line:signing-phase-end}) is run on input
\var{request}.\var{withdrawalData}.\var{signers} as the
signing parties and \var{request}.\var{withdrawalData}.\var{signHash} as the
message. If successful, it returns the signature for the corresponding withdrawal transaction,
in which case the validator updates \var{request}.\var{status} to \const{processed}. If the instance terminates with an error, the validator reverts \var{request}.\var{status} to \const{pending}.
The finalization phase (lines \ref{line:finalization-phase-start} -- \ref{line:finalization-phase-end}) is responsible for submitting the withdrawal transaction to the target chain.

\subsubsection{Consensus phase}\label{sec:consensus-phase}
The goal of the consensus phase is to agree on the request that will be processed, on the transaction hash (\var{signHash}) that needs to be signed (in order to be submitted later, together with the transaction, to the target chain), and on a set of signers that will create the signature (\var{signers}).
This is achieved using two instances of a reliable broadcast (\type{RB}, see \cref{def:reliable-broadcast}) primitive, driven by the proposer.

\begin{algorithm}[ht!]
\caption{Withdrawal-generation protocol, \emph{consensus} phase. Code for the \var{proposer} of session \var{sid}.}
\label{alg:consensus}
\begin{algorithmic}[1]


    \Function{\op{propose()}}{}
        \Let{(\var{depositIdentifier}, \var{request})}{\text{oldest entry in \var{requests} s.t. \var{request}.\var{status} == \const{pending}}}
        \Let{\targetClient}{\text{the \ChainClient for } \var{request.depositData.targetChainID} }
        \Let{\var{signHash}}{\targetClient.\op{getHashOfWithdrawal}(\var{request}.\var{depositData})} \Comment{See Sec.~\ref{sec:chain-clients}} \label{line:evm-proposer-call-signhash}
        \Let{\var{proposalMsg}}{\msgproposal{\var{depositIdentifier}, \var{signHash}}} \label{line:consensus-proposer-form-proposal}
        \State{\type{RB}[\var{sid}, \const{proposal}].\op{broadcast}(\var{proposalMsg})} \label{line:consensus-proposer-broadcast-proposal}
        \State{\textbf{start timer} \var{acceptanceTimeout} \textbf{with duration} \const{BoundaryAcceptance}}
        \Let{\var{possibleSigners}}{\emptyset}

        \Statex
        \On{$\msgacceptance{\var{sid}, \var{depositIdentifier}} \leftarrow \text{validator \var{v}}$} \label{line:consensus-accept-begin}
            \If{\textbf{not} \var{possibleSigners}.\op{contains}(\var{v})}
                \State{{possibleSigners}.\op{insert}(\var{v})}
            \EndIf
        \EndOn\label{line:consensus-accept-end}
        \Statex
        \On{\textbf{timeout} \var{acceptanceTimeout}}\label{line:consensus-end-acceptance}
            \Let{\var{signersCount}}{\op{len}(\var{possibleSigners}) + 1} \Comment{include the proposer}
            \Guard{\var{signersCount} > \const{threshold}}  \label{line:consensus-proposer-require-acceptances} \Comment{otherwise returns \const{error}}
            \Let{\var{signers}}{\textbf{pick } \const{threshold} \textbf{ entries from } \var{possibleSigners} } \Comment{see text for details}
            \State{\var{signers}.\op{insert}(\var{proposer})}
            \Let{\var{signStartMsg}}{\msgsignstart{\var{depositIdentifier}, \var{signers}}} \label{line:consensus-proposer-form-signstart}
            \State{\type{RB}[\var{sid}, \const{signStart}].\op{broadcast}(\var{signStartMsg})} \label{line:consensus-proposer-broadcast-signstart}
            \State{\textbf{return} (\var{depositIdentifier}, \var{signers}, \var{signHash})} \label{line:consensus-proposer-return}
        \EndOn
    \EndFunction
\end{algorithmic}
\end{algorithm}

\begin{algorithm}[ht!]
    \caption{Withdrawal-generation protocol, \emph{consensus} phase. Code for non-proposer validators of session \var{sid}.}
    \label{alg:consensus-validator}
    \begin{algorithmic}[1]
        \Function{\op{accept}()}{}
            \Let{\var{proposedId}}{\bot}
            \Statex

            \On{\type{RB}[\var{sid}, \const{proposal}].\op{deliver}(\var{proposalMsg}, \var{sender}) \textbf{s.t.} \var{sender} == \var{proposer}} \label{line:consensus-receive-proposal}
                \Guard{\var{proposedId} == \bot} \Comment{prevent proposer from sending multiple proposals} \label{line:validator-checks-begin}
                \Let{\var{request}}{\var{requests}(\var{proposalMsg}.\var{depositId})}
                \Guard{\var{request} \neq \bot \textbf{ and } (\var{request}.\var{status} == \const{invalid} \textbf{ or } \var{request}.\var{status} == \const{pending})} \label{line:consensus-acceptor-check-satus}
                \While{$\var{request}.\var{status} == \const{invalid}$}
                    \State{\op{submitWithdrawal}(\var{request})} \label{line:consensus-verify-again} \Comment{Verify the deposit again.}
                \EndWhile
                \Let{\targetClient}{\text{the \ChainClient for } \var{request.depositData.targetChainID}}\label{line:evm-verify-hash-start}
                \Guard{\var{proposalMsg}.\var{signHash} == \targetClient.\op{getHashOfWithdrawal}(\var{request}.\var{depositData})} \label{line:consensus-acceptor-check-hash} \label{line:evm-verify-hash-end} \label{line:validator-checks-end}
                \State{$\msgacceptance{\var{sid}, \var{proposalMsg}.\var{depositId}} \rightarrow \var{proposer}$}
                \Let{\var{proposedId}}{\var{proposalMsg}.\var{depositId}}
            \EndOn
            \Statex

            \On{\type{RB}[\var{sid}, \const{signStart}].\op{deliver}(\var{signStartMsg}, \var{sender}) \textbf{s.t.} \var{sender} == \var{proposer}} \label{line:consensus-receive-signstart}
                \Guard{\var{proposedId} == \var{signStartMsg}.\var{depositId}}\label{line:consensus-accepted-deposit}
                \Comment{Ignore if proposal differs or not delivered}
                \State{\textbf{return} (\var{depositId}, \var{signStartMsg}.\var{signers}, \var{proposalMsg}.\var{signHash})} \label{line:consensus-validator-return}
            \EndOn
        \EndFunction
    \end{algorithmic}
    \end{algorithm}

Specifically, as shown in \Cref{alg:consensus}, the proposer picks the oldest request with a \const{pending} status from its local database
and uses the \ChainClient for the target chain to compute the hash of its \var{depositData} (\cref{line:evm-proposer-call-signhash}).
The proposer forms a \msgproposalEMPTY{} message containing the \var{depositId} and the \var{signHash} (\cref{line:consensus-proposer-form-proposal}) and broadcasts it to all validators (\cref{line:consensus-proposer-broadcast-proposal}) using a reliable broadcast primitive.

Upon delivering this proposal (\cref{line:consensus-receive-proposal} of \Cref{alg:consensus-validator}), a validator retrieves the corresponding request from its local memory. It is required that the request is not being processed, hence its status is either \const{invalid} or \const{pending}.
If its status is \const{invalid}, it tries to verify the request again (\cref{line:consensus-verify-again}).
If the request indeed becomes \const{pending} (which means the deposit transaction has been finalized on the source chain),
the validator verifies the \var{signHash} sent by the proposer (lines \ref{line:evm-verify-hash-start}--\ref{line:evm-verify-hash-end}).
If all checks out, it responds with an \msgacceptanceEMPTY{} message.

The proposer keeps track of the validators that have sent an \msgacceptanceEMPTY{} message (lines \ref{line:consensus-accept-begin}--\ref{line:consensus-accept-end} of \Cref{alg:consensus}).
It stops handling \msgacceptanceEMPTY{} messages after a predefined duration (\cref{line:consensus-end-acceptance}).
At that point, the proposer (deterministically, see below for more comments) selects $t + 1$ validators (always including itself). It creates a \msgsignstartEMPTY{} message containing the \var{signers} of the session (\cref{line:consensus-proposer-form-signstart}) and broadcasts it to all \var{signers} using a reliable broadcast primitive (\cref{line:consensus-proposer-broadcast-signstart}).
Upon receiving this message (\cref{line:consensus-receive-signstart} of \Cref{alg:consensus-validator}), a validator stores \var{signers} as the set of signers for the request indicated by \var{\var{depositId}}. Finally, the proposer and all validators mark the status of the request as \const{processing}.

We note here that this logic is agnostic to the target chain. The \var{signHash} will always be the hash of the withdrawal transaction for the target chain. The details for computing and verifying the \var{signHash} are chain-specific, and are delegated to the \ChainClient for the target chain.

\begin{remark}[Verifying the validity of the deposit transaction in a loop]
    The reason for verifying the validity of the deposit transaction again is that the validator may have received the \op{submitWithdrawal()} call before the deposit transaction was finalized on the source chain, hence leaving the request in \const{invalid} status, while the proposer received it later and marked it as \const{pending}. As we see in the analysis section, the duration of the acceptance sub-phase must be sufficiently long, so that all validators have time to see the deposit transaction as finalized. We note that the logic of continuously trying to verify a request for some time interval can be implemented elsewhere in the protocol, even in an asynchronous manner.
\end{remark}

\begin{remark}[Choosing the signers set]
The proposer uses a pseudo-random generator (PRG), seeded with session id \var{sid}, in order to select $t$ signers out of the $m$ acceptors (validators that have accepted the proposal). More precisely, this is done as follows: The acceptors are first sorted based on their address. The PRG is used to obtain a 64-bit number, which is then reduced modulo $m$, pointing to an acceptor. That acceptor is removed, $m$ is decremented, and the process is repeated until $m = t$.

We observe that the only randomness in this process is the set of validators that accept the proposal. Given this set, it is a \emph{deterministic} process, as is \var{sid} is known to all validators -- any validator can run the same algorithm and predict the set of signers.
Moreover, the validators do \emph{not} verify the set of signers in the \msgsignstartEMPTY message, which means the proposer does not have to follow the prescribed algorithm to choose \var{signers}, but it can choose any $t$ acceptors.

This implies that the proposer has complete control over the set of signers, and can choose it to be any set of $t$ validators that have accepted the proposal.
\end{remark}

\subsubsection{Signing phase}\label{sec:signing-phase}

\begin{algorithm}[ht!]
    \caption{Withdrawal-generation protocol, \emph{signing} phase. Code for a validator \var{v} of session \var{sid}, parameterized by the bridge's public key $\var{pk}$.}
    \label{alg:signing-phase}
    \begin{algorithmic}[1]
        \Function{\op{sign}(\var{signHash}, \var{signers})}{}
            \If{$\var{v} \in \var{signers}$}
                \Let{\var{result}}{\text{run binance-TSS with message \var{singHash} and parties \var{signers}}} \label{line:signing-run-tss}
                \State{\type{RB}[\var{sid}, \const{signature}].\op{broadcast}((\var{result}, \var{signHash}))} \label{line:broadcast-signature}
            \EndIf
            \Statex
            \On{\type{RB}[\var{sid}, \const{signature}].\op{deliver}((\var{result}, \var{signHash}), \var{sender}) \textbf{s.t.} $\var{sender} \in \var{signers}$} \label{line:signing-receive-signature}
                \If{$\var{result} \neq \const{error}$}
                    \Let{\var{signature}}{\var{result}}
                    \If{\text{binance-TSS.\op{verify}(\var{pk}, \var{signHash}, \var{signature})}}
                        \State{\textbf{return} \var{signature}} \label{line:signing-verify-signature}
                    \EndIf
                \EndIf
        \EndOn
        \EndFunction
    \end{algorithmic}
    \end{algorithm}

The second phase of the withdrawal generation is the \emph {signing phase}, shown \Cref{alg:signing-phase}.
A validator runs the threshold-signing protocol only if included in the \var{signers} variable of the consensus phase (\cref{line:signing-run-tss}).
The signers run an instance of the signing protocol of the \emph{binance-TSS}
library (\cref{def:binance-tss}). If the instance terminates successfully, the signers obtains
\var{signature}, which the broadcast among all validators.
A validator ends the signing phase if it delivers a signature that verifies for message \var{signHash} (\cref{line:signing-verify-signature}).

\begin{remark}[The size of the signers set]
    The signing protocol used by Bridgeless is the ECDSA threshold-signature protocol of Gennaro and Goldfeder~\cite{DBLP:journals/iacr/GennaroG19}, implemented in the bnb-chain repository\footnote{\url{https://github.com/bnb-chain/tss-lib/tree/v2.0.2}}. The signing phase of that protocol will \emph{abort} if any signer misbehaves, and it must consequently be restarted, potentially on a different set of signers. This is because the protocol works in the so-called \emph{dishonest majority} setting, which means it tolerates any threshold $t \leq n$.
    As a result, a single malicious validator can block the completion of a bridging request.

    The protocol of \btm mitigates this issue in two ways. First, it chooses the smallest possible signers set in the consensus phase, even if more validators have sent an acceptance message. This also keeps the communication among the signers as low as possible. Second, the protocol detects and punishes certain malicious actions (but this is done manually and outside the main bridging protocol, and is not implemented at the time of writing). In \Cref{sec:analysis} we remark that these measures are not sufficient, when the actual number of malicious parties is close to $t$.
\end{remark}

\subsubsection{Finalization phase}\label{sec:finalization-phase}
At this point, the required signature for the withdrawal transaction has been computed and locally stored by all signers of the current session. The goal of the finalization phase is to submit the withdrawal transaction to the target chain. The \op{finalize()} function is shown in \Cref{alg:finalization-phase}.

\begin{algorithm}[ht!]
    \caption{Withdrawal-generation protocol, \emph{finalization} phase. Code for a validator \var{v} of session \var{sid}.}
    \label{alg:finalization-phase}
    \begin{algorithmic}[1]
        \Function{\op{finalize}(\var{depositId}, \var{depositData}, \var{signature})}{}
            \If{\var{requests}[\var{depositId}].\var{status} $\neq$ \const{processed}}
                \State{\textbf{return} $\bot$}
            \EndIf
            \Let{\targetClient}{\text{the \ChainClient for } \var{depositData}.\var{targetChainID}}
            \If{\var{depositData}.\var{targetChain} == \const{EVM}}
                \State{\targetClient.\op{submitTx}(\var{depositId}, \var{depositData}, \var{signature})}
            \ElsIf{\var{depositData}.\var{targetChain} == \const{Zano}}
                \Let{\var{tx}}{\targetClient.\op{getWithdrawalTx}(\var{depositId}, \var{depositData})}
                \State{\targetClient.\op{submitTx}(\var{tx}, \var{signature})}
            \ElsIf{\var{depositData}.\var{targetChain} == \const{Bitcoin}}
                \Let{\var{tx}}{\targetClient.\op{getWithdrawalTx}(\var{depositId}, \var{depositData})}
                \Let{\var{signedTx}}{\op{injectSignatures}(\var{tx}, \var{signature})} \Comment{details omitted}
                \State{\targetClient.\op{submitTx}(\var{signedTx})}
                \Let{\var{lockedBitcoinInputs}}{\var{lockedBitcoinInputs} \cup \var{tx}.\var{in}}
            \EndIf
            \State{\textbf{return} \const{true}}
        \EndFunction
    \end{algorithmic}
    \end{algorithm}

Function \op{finalize()} is run by all validators. Depending on the target chain, the validator does slightly different things:

\dotparagraph{EVM}
If the target chain is an EVM chain, then the validator can use the \op{submitTx()} function of the EVM \ChainClient (\Cref{alg:evm-chain-client}), which will call the corresponding smart-contract function. The protocol also allows clients to submit the withdrawal transaction to the target chain. We describe this in \Cref{sec:withdrawal}.

\dotparagraph{Zano}
If the target chain (according to \var{request}.\var{depositData}.\var{targetChain}) is Zano, the validator constructs the withdrawal transaction using functions \op{getWithdrawalTx()} and submits it, together with the signature, to the Zano network using \op{submitTx()} of the \ChainClient for Zano (\Cref{alg:zano-chain-client}).

\dotparagraph{Bitcoin}
At this point, the validator holds the \var{signature} variable for the withdrawal transaction, which may internally contain more than one one signatures -- one for each input of the bitcoin transaction. The validator then uses function \op{injectSignatures}() to update the bitcoin transaction with the signatures -- we omit the details, but the idea is to update the unlocking script (witness) of the inputs of -- and submits it to the bitcoin network using a \ChainClient for bitcoin (\Cref{alg:btc-chain-client}).
Finally, the validator updates the \var{lockedBitcoinInputs} variable of its local state (defined in \Cref{alg:data-types}), which holds all UTXOs that have been used in bitcoin withdrawal transactions. The bitcoin chain client takes this into account when collecting inputs for a new transaction.

The function returns \const{true} if the transaction was successfully submitted, and an $\error$ otherwise.
\section{Analysis}\label{sec:analysis}

This section covers the security analysis of the bridgeless protocol
(\cref{sec:bridge-protocol}). As outlined above, the core properties that the
protocol should guarantee are bridge liveness and safety
(\cref{def:bridge-liveness} and \ref{def:bridge-safety}). Next, we distill the
properties that each phase of the protocol should satisfy via an array of
lemmas. Eventually, by using the lemmas as stepping stones, we show that the
protocol satisfies the outlined properties.

In the following analysis we assume that:
\begin{itemize}
    \item the network is \emph{synchronous}, s.t. all messages are delivered within a (known) time bound $\Delta$, and the durations of each phase (\const{acceptanceBoundary}, \const{consensusBoundary}, \const{signBoundary}, \const{finalizationBoundary}) are long enough for each phase to complete.
    \item at most $t$ out of all $n$ validators are adversarial and $n \geq 3t + 1$
    \item every honest validator runs a \emph{honest chain client} on all source and target chains
    \item all source and target chains are secure, i.e., guarantee ledger safety and liveness
\end{itemize}

\begin{remark}[Bulk transaction processing]
    The following analysis assumes the existence of a single request.
    Intuitively, this simplification corresponds to the oldest non-processed
    request in the system, which is the one that honest proposers
    deterministically choose to propose. This simplification is needed in order
    to get meaningful liveness guarantees, because the bridge can only process
    one request (per source chain) at a time. Therefore, the starting point for
    evaluating liveness is when a transaction becomes eligible to be chosen by
    an honest validator, abstracting previous delays e.g., due to high traffic
    and the submission of multiple requests in parallel.
\end{remark}

\subsection{Chain clients}
We now state and prove some lemmas about the \op{getDepositData}() function of a \ChainClient.

\begin{lemma}[Chain-client liveness]\label{lem:chain-client-liveness}
    Function \op{getDepositData}(), called by an honest validator with a \var{depositId} generated by an honest client, will eventually return a value different from \const{error}.
\end{lemma}
\begin{proof}
    The proof follows from the liveness of the underlying ledger and from the fact that an honest client creates the deposit transaction (\Cref{sec:deposit}) in the format expected by the chain clients (\Cref{sec:chain-clients}).
\end{proof}

\begin{lemma}[Chain-client agreement]\label{lem:chain-client-agreement}
    If two honest validators call function \op{getDepositData}() with the same \var{depositId} and the calls do not return \const{error}, then they return the same value (of type \type{DepositData}) to both clients.
\end{lemma}
\begin{proof}
    The proof follows from the assumption that each underlying source chain is safe, hence, if the deposit transaction is found in it, then it will always be the same transaction. Specifically:

    The EVM chain client (\cref{alg:evm-chain-client}) uses function \op{eth\_getTransactionByHash} of the RPC provider to retrieve the deposit transaction. Assuming the RPC provider is honest and that two transactions do not have the same hash (which is true with all but negligible probability), this returns either an error (if \var{txHash} is not found) or the same \var{tx} on each call. Similarly for the call to \op{eth\_getTransactionReceipt}, which returns \var{receipt}. All fields of \type{DepositData} are retrieved from \var{tx} and \var{receipt}, hence, they are the same.
    For the Zano chain client (\cref{alg:zano-chain-client}) and the bitcoin chain client (\cref{alg:btc-chain-client}), similar reasoning applies, this time without trusting an RPC provider, as each \btm validator runs a Zano full node and a bitcoin full node.
\end{proof}

\begin{lemma}[Chain-client safety]\label{lem:chain-client-safety}
    If function \op{getDepositData}() returns some \type{DepositData}, then the corresponding deposit transaction has been finalized on the source chain.
\end{lemma}
\begin{proof}
    The proof follows from the safety of the underlying ledger and from the fact that the chain clients require the necessary number of confirmations to be found in the source chain.
\end{proof}

\subsection{Deposit-verification protocol}
The deposit-verification subprotocol ensures the validity of bridging requests.
\cref{lem:deposit-verification-agreement} formalizes the requirement that all honest
validators agree on the validity of a request. This result will be useful
to eventually prove bridge liveness, wherein an honestly-created request is eventually
processed by the bridge.

\begin{lemma}[Deposit-verification liveness]\label{lem:deposit-verification-liveness}
    If an honest client calls \op{submitWithdrawal}() on some \var{depositId}, then all honest validators will eventually mark the corresponding request as \const{pending}.
\end{lemma}
\begin{proof}
    From \Cref{lem:chain-client-liveness}, the call in \cref{line:evm-call-get-deposit} will eventually return some \var{depositData}. Since an honest client creates a well-formed \var{depositId}, the rest of the checks of \op{submitWithdrawal}() will pass and an honest validator will mark the request as \const{pending}.
\end{proof}

\begin{lemma}[Deposit-verification agreement]\label{lem:deposit-verification-agreement}
    (1) If an honest validator marks a request as \const{pending}, then all honest
    validators eventually mark it as \const{pending}.
    (2) If two honest validators mark a request as \const{pending}, they hold the same \var{depositData} for that request.
\end{lemma}
\begin{proof}
    (1) The proof follows from two conditions of \cref{alg:tss}.
    First, all honest validators will receive a \msgsubmitwithrawal{} for the
    given request. This is guaranteed by \cref{line:evm-echo-withdrawal-begin},
    where an honest validator directly sends this message to all other
    validators.
    Second, all conditions of function $\op{submitWithdrawal}$ will eventually be
    satisfied for all honest validators. This follows from \Cref{lem:chain-client-liveness}. Note that an honest validator may not immediately observe a valid deposit transaction as finalized, but it will eventually do so after the liveness parameter of the source chain has passed.
    (2) The variable \var{depositData} is constructed in \cref{line:evm-call-get-deposit} of \Cref{alg:tss} using \op{getDepositData}(). The result follows from \cref{lem:chain-client-agreement}.
\end{proof}

\begin{lemma}[Deposit-verification safety]\label{lem:deposit-verification-safety}
    If an honest validator marks a request as \const{pending}, then the corresponding deposit transaction has been finalized on the source chain.
\end{lemma}
\begin{proof}
    The proof follows from \Cref{lem:chain-client-safety}.
\end{proof}

\subsection{Withdrawal-generation protocol}

At a high level, for the withdrawal generation subprotocol we will need to prove
the following two properties. First, we will show that all requests that
correspond to a valid deposit, i.e., which are marked as \const{pending} by honest
validators, are eventually processed and finalized (liveness). Second, we will
show that any finalized request has been marked as \const{pending} by an honest
validator (safety). To prove these two properties we first outline
intermediary properties via lemmas that correspond to each phase of the
withdrawal generation.

\subsubsection{Consensus phase}\label{sec:consensus-phase-analysis}

During the consensus phase, a designated party (proposer) chooses a request and
a subset of validators which, including itself, form the signing
committee for that request.

\begin{lemma}[Consensus liveness]\label{lem:consensus-liveness}
    If the proposer is honest
    and all honest validators start the session with the request chosen by the proposer as \const{pending},
    then at the end of the consensus phase all honest
    validators will change some request's status from \const{pending} to
    \const{processing}.
\end{lemma}
\begin{proof}
    The honest proposer picks the request with status \const{pending} and correctly computes its \var{signHash}
    and broadcasts a \msgproposalEMPTY{} message (\cref{line:consensus-proposer-broadcast-proposal} of \Cref{alg:consensus}).
    By the properties of reliable broadcast, all honest validators deliver it (\cref{line:consensus-receive-proposal} of \Cref{alg:consensus-validator}).
    By the assumption that honest validators start the session with the request chosen by the proposer as \const{pending}, the checks in lines \ref{line:consensus-acceptor-check-satus} and \ref{line:consensus-acceptor-check-hash} pass, and honest validators send an \msgacceptanceEMPTY message to the proposer.
    By assumption, there are at least $n-t \geq 2 \cdot t + 1$ honest validators,
    we are in synchrony and \const{acceptanceBoundary} is long enough,
    the condition on \cref{line:consensus-proposer-require-acceptances} of \Cref{alg:consensus} will be satisfied.
    The honest proposer fills the set \var{signers} with $\const{threshold} + 1 = 2t + 1$ party identifiers (including itself),
    creates a \msgsignstartEMPTY containing \var{signers} and reliably-broadcasts it (\cref{line:consensus-proposer-broadcast-signstart}).
    Following, all honest validators deliver it (\cref{line:consensus-receive-signstart} of \Cref{alg:consensus-validator}).
    Eventually, in the end of the consensus phase, all honest validators store the same \var{signHash} and \var{signers} values for the request.
    The proposer returns the request's identifier in \cref{line:consensus-proposer-return} of \Cref{alg:consensus} and the rest of the validators in \cref{line:consensus-validator-return} of \Cref{alg:consensus-validator}.
    Observe that, since \const{consensusBoundary} is long enough, all these will happen before the consensus timeout expires.
\end{proof}

\begin{lemma}[Consensus agreement]\label{lem:consensus-agreement}
    (1) If an honest validator marks a request as \const{processing}, then all honest validators that started the consensus phase with that request as \const{pending} will mark it as \const{processing}.
    (2) If two honest validators change some request's status to \const{processing}, then they hold the same \var{signHash} and \var{signers} for that request.
\end{lemma}
\begin{proof}
    (1) From the \emph{agreement} property of reliable broadcast (\Cref{def:reliable-broadcast}), if one honest validator delivers the \msgproposalEMPTY{} message on \cref{line:consensus-receive-proposal} of \Cref{alg:consensus-validator}, then all honest validators deliver the same message.
    From \Cref{lem:deposit-verification-agreement}, honest validators hold the same \var{depositData}.
    Hence, the checks in Lines~\ref{line:validator-checks-begin}--\ref{line:validator-checks-end} will pass and they all send an \msgacceptanceEMPTY message. Similarly, if one honest validator delivers the \msgsignstartEMPTY{} message on \cref{line:consensus-receive-signstart} of \Cref{alg:consensus-validator}, then all honest validators deliver the same message, and they return the same \var{depositId} on \Cref{line:consensus-validator-return}.
    (2) Continuing from (1), honest validators return the same data on \Cref{line:consensus-validator-return}, hence \Cref{alg:withdrawal-generation} on \Cref{line:obtain-consensus-result} for honest validators receives the same results.
\end{proof}

\begin{lemma}[Consensus safety]\label{lem:consensus-safety}
    If an honest validator marks a request as \const{processing}, then the validator has previously marked the request as \const{pending}.
\end{lemma}
\begin{proof}
    This follows from \cref{alg:consensus-validator} from: (i)
    \cref{line:consensus-acceptor-check-satus} that requires a validator to
    reply with an acceptance message only if the request's status is pending;
    (ii) \cref{line:consensus-accepted-deposit} that requires a validator to
    change a request's status to \const{pending} only if it has previously sent
    an acceptance message.
\end{proof}

\subsubsection{Signing phase}\label{sec:signing-phase-analysis}

After the consensus phase ends, a request has been chosen and a committee
formed to process it. The following lemmas show that a request is processed, and
a signature for it is generated, as long as at least one honest validator
processes it (necessary condition) and certainly if both the request and all
committee members are honest (sufficient condition).

\begin{lemma}[Signature-generation liveness]\label{lem:signature-liveness}
    If all parties in the set \var{signers} are honest, then all honest validators
    will set the request's status to \const{processed} at the end of the
    signing phase.
\end{lemma}
\begin{proof}
    From the liveness property of TSS
    (\cref{def:binance-tss}), wherein if all signers are honest then a valid
    signature is produced, we get that \cref{line:signing-run-tss} of \Cref{alg:signing-phase} will return a valid signature,
    which will be then broadcast by honest signers (\cref{line:broadcast-signature}), delivered by all honest validators (\cref{line:signing-receive-signature}), and successfully verified by all honest validators (\cref{line:signing-verify-signature}).
    Therefore, \Cref{alg:signing-phase} will return a valid signature and line \ref{line:mark-processed} of \Cref{alg:withdrawal-generation} will be executed.
\end{proof}

\begin{lemma}[Signature-generation agreement]\label{lem:signature-agreement}\label{lem:signature-error-agreement}
    (1) If an honest validator marks a request as \const{processed}, then all
    honest validators
    will mark the request as \const{processed} at the end of the signing phase.
    (2) If an honest validator reverts the request to \const{pending} at the end of
    the signing phase, then all honest validators do the same.
\end{lemma}
\begin{proof}
    (1) If an honest validator marks a request as processed, then it has received a valid signature on the \var{signHash} of the request.
    This signature has been broadcast to all validators (\cref{line:broadcast-signature} of \cref{alg:withdrawal-generation}).
    Therefore, by the properties of the broadcast protocol, all honest
    validators receive the signature and also mark the request as processed.
    Observe that, since we are in synchrony and \var{signBoundary} is long enough, honest validators will have received the signature before the signing timeout expires (\cref{line:signing-phase-timeout-end} of \Cref{alg:withdrawal-generation}).
    (2) From the agreement property of the TSS protocol (\cref{def:binance-tss}), all honest signers receive the same signature result. If the result is \const{error}, then no validator will be able to broadcast a valid signature, except with negligible probability. Hence, all honest validators will timeout (line \ref{line:signing-phase-timeout-end} of \Cref{alg:withdrawal-generation}) with $\var{signResult} == \bot$ and mark the request as \const{pending}.
\end{proof}

We note here that the protocol allows a validator $v$ to mark a request as \const{processed} even if it has not previously marked the request as \const{processing} -- this can happen, for example, if the proposer and $2t$ honest validators make progress in the consensus and signing phases but $v$ has not yet observed the deposit transaction as finalized, or if $v$ goes offline. The next lemma ensures that if a validator such as $v$ sets the request's status to \const{processed}, then at least $t+1$ honest validators have previously marked the request as \const{processing}.

\begin{lemma}[Signature-generation safety]\label{lem:signature-safety}
    If an honest validator marks a request as \const{processed}, then at least $t+1$ honest validators have previously marked the request as \const{processing}.
\end{lemma}
\begin{proof}
    In \cref{line:signing-phase-end} \cref{alg:withdrawal-generation}, an
    honest validator $v$ marks the request as \const{processed} only if the TSS
    instance terminates without an error and produces a valid signature on the \var{signHash} of the request. For this to
    happen, all parties in the signing committee should participate in signing
    the request. Since the size of the signing committee is $2t+1$, this happens
    if at least $t+1$ honest validators sign the request.
    From \cref{alg:consensus-validator}, these validators have
    run \cref{line:mark-processing} of \Cref{alg:withdrawal-generation} and marked the request as \const{processing},
    otherwise they would have abandoned the session on \cref{line:obtain-consensus-result}.
\end{proof}

\subsubsection{Finalization phase}\label{sec:finalization-phase-analysis}

The end phase of the withdrawal subprotocol is finalization. At this point, a
signature for a request has been generated, at the end of the signing phase,
and the withdrawal transaction that contains it needs to be published on the
target chain. The following lemmas show that, if such signature exists
then it will be published at the destination (\cref{lem:finalization-liveness})
and, if a request is finalized then such signature necessarily exists for it
(\cref{lem:finalization-safety}), such that all honest validators mark it as
finalized (\cref{lem:finalization-agreement}).

\begin{lemma}[Finalization liveness]\label{lem:finalization-liveness}
    If an honest party has marked a request as \const{processed}, then all
    honest parties will mark the request as \const{finalized} in the end of the
    finalization phase.
\end{lemma}
\begin{proof}
    If an honest party marks the request as processed, then they will create the
    final transaction to the destination chain and submit it to the client
    (\cref{alg:finalization-phase}), after which point they mark the request as
    finalized. Also, from \cref{lem:signature-agreement}, if one honest party
    had marked the request as processed, then all other parties do so, hence
    they will all mark it as finalized as well.
\end{proof}

\begin{lemma}[Finalization agreement]\label{lem:finalization-agreement}
    If an honest party marks a request as \const{finalized}, then all honest
    parties will mark the request as \const{finalized} at the end of the
    finalization phase.
\end{lemma}
\begin{proof}
    This follows from the observation that the transaction that all honest
    parties submit to the target ledger during the finalization phase
    (\cref{alg:finalization-phase}) is the same. Therefore, if an honest party
    marks it as finalized, then this party has received a publication receipt
    from the target client. Since the target ledger satisfies safety by
    assumption, all honest parties will receive a receipt, so they will also
    mark the request as finalized.
\end{proof}

\begin{lemma}[Finalization safety]\label{lem:finalization-safety}
    If an honest party marks a request as \const{finalized}, then the party has
    marked this request as \const{processed}.
\end{lemma}
\begin{proof}
    This follows from the following observations. First, a request is marked as
    processed if a valid signature is produced (\cref{lem:signature-safety}), in
    which case all honest parties receive it (\cref{lem:signature-agreement}).
    Second, the party submits the request to the target client only if it has
    been marked as processed. When the signature is published on the ledger, the
    party marks the request as finalized. This happens if the target ledger
    satisfies liveness (which it does by assumption) and the signature is valid,
    which is a necessary condition for the request to have been marked as
    processed.
\end{proof}

\subsubsection{Withdrawal security}

We are now ready to prove that the withdrawal generation subprotocol satisfies
the two properties described above.

\begin{figure}
    \center
    \includegraphics[width=0.9\columnwidth]{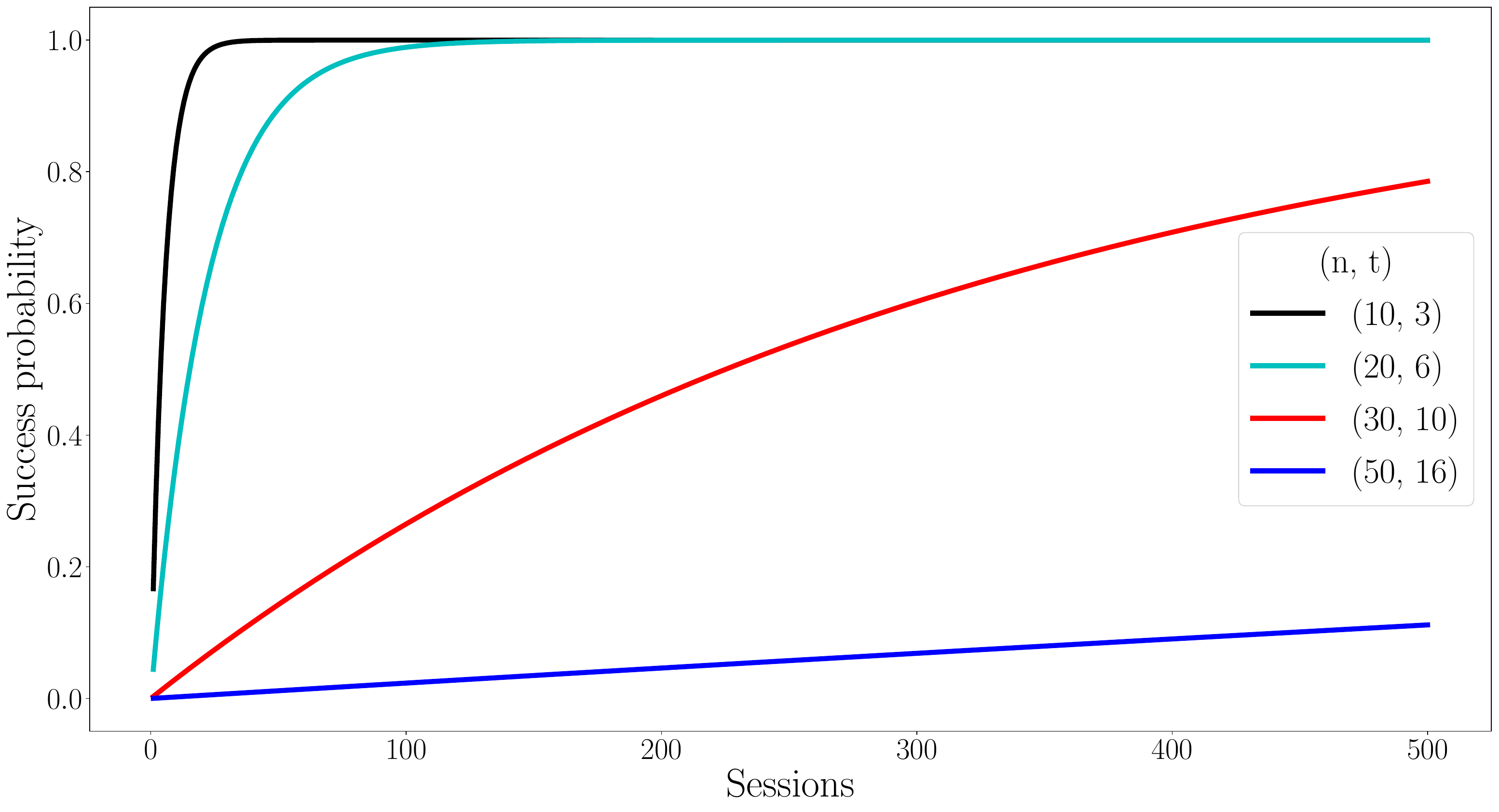}
    \caption{
        Estimation of the probability of success in finalizing a request
        (y-axis) w.r.t. the number of executed sessions (x-axis) for various
        values of $n$ (total parties), where $t = \frac{n}{3}$ (adversarial
        parties) (cf. \cref{thm:withdrawal-liveness}).
    }
    \label{fig:liveness-simulation}
\end{figure}

First, \cref{thm:withdrawal-liveness} proves that all pending requests are
eventually finalized (liveness). Additionally, \cref{fig:liveness-simulation}

\begin{theorem}[Withdrawal liveness]\label{thm:withdrawal-liveness}
    If a request is marked as \const{pending} by all honest validators, then it
    will be marked as \const{finalized} by all honest validators after
    $s$ sessions with probability
    $1 - (1 - \frac{n-t}{n} \cdot \prod_{i=1}^{t} \frac{n-i-t}{n-i})^r$.
\end{theorem}
\begin{proof}
    The proof follows directly from the observation that, if all signing
    committee members are honest, then the request's status will change for all
    honest validators from \const{pending} to \const{processing} to
    \const{processed} to \const{finalized}. This comes from the liveness lemmas
    of each phase of the withdrawal subprotocol (Lemmas
    \ref{lem:consensus-liveness},
    \ref{lem:signature-liveness},
    \ref{lem:finalization-liveness}).
    Additionally, if one of the signing committee members is faulty and the
    signing fails, then all honest validators mark the request as pending
    (\cref{lem:signature-error-agreement}).

    Therefore, it suffices to compute the probability that the proposer and all
    other members of the signing committee (which includes the proposer) are all
    honest when chosen at random.

    Choosing an honest proposer at random occurs with probability
    $p_\text{hprop} = \frac{n-t}{n}$.

    The probability that an honest proposer chooses an all-honest signing
    committee of $t+1$ members is at least:
    $$p_\text{hcom} = \prod_{i=1}^{t} \frac{n-i-t}{n-i}$$
    Note that, since all honest validators respond with an \msgacceptanceEMPTY
    to the proposer during the consensus phase, then this probability comes up
    when all adversarial parties also respond with \msgacceptanceEMPTY. In this
    case, the proposer chooses the committee members at random from the set of
    all parties.

    Therefore, the probability that, in a given iteration of the protocol, all
    signing committee members, that is both the proposer and the chosen
    parties, are all honest is:
    $p_\text{h} = \frac{n-t}{n} \cdot \prod_{i=1}^{t} \frac{n-i-t}{n-i}$.

    Consequently, after $r$ rounds, the probability that an all-honest signing
    committee has been elected is at least:
    $$p_\text{liveness} = 1 - (1 - \frac{n-t}{n} \cdot \prod_{i=1}^{t} \frac{n-i-t}{n-i})^r$$
\end{proof}

In \Cref{fig:liveness-simulation} we pictorially show this probability for various values of $n$ and $t$. Each line shows the probability to finalize one request after a number of sessions of the withdrawal generation protocol.

Next, \cref{thm:withdrawal-safety} shows that all finalized requests have been
previously marked as pending by honest validators (safety).

\begin{theorem}[Withdrawal safety]\label{thm:withdrawal-safety}
    If an honest validator marks a request as \const{finalized}, then at least
    one honest validator has marked it as \const{pending}.
\end{theorem}
\begin{proof}
    The proof follows from the safety and agreement lemmas of each phase of the
    subprotocol. Specifically, a honest validator changes the request's status
    to \const{processing} only if they have marked it as \const{pending}
    (\cref{lem:consensus-safety}), followed by changing it to \const{processed}
    (\cref{lem:signature-safety}), followed by changing it to \const{finalized}
    (\cref{lem:finalization-safety}). Additionally, due to the agreement lemmas
    of each phase (\cref{lem:consensus-agreement},
    \cref{lem:signature-agreement}, \cref{lem:finalization-agreement}), if one
    honest validator proceeds to change the request's status from
    \const{pending} to \const{processing} to \const{processed} to
    \const{finalized}, then all honest validators do the same sequence of status
    changes.
\end{proof}

\subsection{Bridge security}

At this point, our protocol's security w.r.t. the original properties of a
bridge's definition is a direct result of previous lemmas.

\begin{theorem}[Bridgeless Liveness]\label{thm:liveness}
    The protocol of \cref{sec:bridge-protocol} satisfies bridge liveness
    (\cref{def:bridge-liveness}) with parameter $s \cdot \lambda$, where
    $\lambda$ is the time duration of a session, with probability
    $1 - (1 - \frac{n-t}{n} \cdot \prod_{i=1}^{t} \frac{n-i-t}{n-i})^r$.
\end{theorem}
\begin{proof}
    The proof follows from \Cref{lem:deposit-verification-liveness}, which shows that, if an honest client performs a deposit, then all honest validators eventually mark the corresponding request as \const{pending}, and from \Cref{thm:withdrawal-liveness}, which shows that, if a request is marked as \const{pending} by all honest validators, it will eventually become \const{finalized} -- which, by definition of the \const{finalized} state, means that the withdrwal has been finalized on the target chain.
\end{proof}

\begin{theorem}[Bridgeless Safety]\label{thm:safety}
    The protocol of \cref{sec:bridge-protocol} satisfies bridge safety
    (\cref{def:bridge-safety}) with overwhelming probability.
\end{theorem}
\begin{proof}
    The proof follows from \Cref{lem:deposit-verification-safety} and \Cref{thm:withdrawal-safety}.
\end{proof}

\subsection{Better liveness through accountability}
We have seen that a \emph{single} malicious validator can attack the liveness of the protocol, if it ends up in the \var{signers} set but does not send a signature on the withdrawal transaction. The protocol mitigates this risk by choosing the smallest possible \var{signers} set. However, in particular for large values of $n$ and $t$, this still leads to a small liveness parameter, as shown in \Cref{fig:liveness-simulation}. This is because with high probability at least one malicious validator will be in \var{signers}.

In such a case, the only way to improve liveness is by detecting the misbehaving validators, i.e., by adding \emph{accountability} to the protocol. This requires several additions. First, the threshold-signing protocol used between the \var{signers} must satisfy and \emph{identifiable-abort} property, meaning that, for every instance where the protocol is unable to output a valid signature, it identifies at least one misbehaving signer. Second, the protocol of Bridgeless must be able to use these identifying messages in order to punish (or altogether remove from the set of validators) the misbehaving signers. At the time of writing this report, this identification is done manually outside of the main protocol. For this reason, the accountability is left out of scope of this analysis.
Third, the validators (after removing the misbehaving ones) must agree on the status of the requests in the system, even if some of the previous sessions have failed, so as to process all remaining requests again. The agreement lemmas in this analysis (Lemmas \ref{lem:deposit-verification-agreement}, \ref{lem:consensus-agreement}, \ref{lem:signature-agreement}, \ref{lem:finalization-agreement}) describe the conditions under which this happens.

\section{Deposit Implementation}\label{sec:deposit}
The deposit is initiated by the client. At this end of this procedure the client constructs a variable \var{depositInfo}, containing the following:
\begin{description}
    \item [\var{txHash}:] It is the hash of the deposit transaction, i.e., the transaction that contains the deposit operation
    \item [\var{txNonce}:] It is an \emph{index}, interpreted differently for each source chain: for EVM chains it is the index of the event emitted by the deposit transaction, for bitcoin it is an index of an output UTXO of the deposit transaction, and for Zano it is an index to the \var{serviceEntries} array of the deposit transaction.
    \item [\var{chainID}:] It is the identifier of the source chain.
\end{description}

\subsection{Deposit on an EVM chain}
For depositing on an EVM-compatible chain, the client calls a smart contract, deployed by \btm on the source chain. Subsequently, the client creates a \var{depositIdentifier}, which it forwards to the validators of \btm.

\dotparagraph{Smart contracts}
In order to support EVM chains as source (i.e., depositing on EVM chains), a number of smart contracts are deployed to that chain, shown in \Cref{alg:evm-deposit-contracts}. These allow clients to deposit native source-chain tokens or arbitrary ERC20 tokens.
\begin{algorithm}[ht!]
\caption{Smart contracts for EVM source chains}
\label{alg:evm-deposit-contracts}
\begin{algorithmic}[1]
    \Function{\op{depositErc20}(}{}
    \State{     \var{token}}, \Comment{address of token to be transferred}
    \State{     \var{amount}}, \Comment{amount of tokens to transfer}
    \State{    \var{receiver}}, \Comment{receiver address on the target network}
    \State{    \var{network}}, \Comment{target network identifier}
    \State{)}
    \Guard{amount > 0}    \Comment{\var{msg}: the tx that called the contract}
    \State{\type{erc20}.\op{transferFrom}(\var{msg.sender}, \var{this}, \var{amount})} \Comment{\var{this}: the address of the contract}
    \State{\textbf{emit} \event{DepositedErc20}\var{(token, amount, receiver, network\varIsWrapped)}}
    \EndFunction
    \Statex

    \Function{\op{depositNative}(}{}
        \State{    \type{string} \var{receiver},} \Comment{receiver address on the target network}
        \State{    \type{string} \var{network}} \Comment{target network identifier}
        \State{)}
        \Guard{\var{msg}.\var{value} > 0}
        \Let{\var{amount}}{\var{msg}.\var{value}}
    \State{\textbf{emit} \event{DepositedNative}(\var{amount, receiver, network})}
    \EndFunction
\end{algorithmic}
\end{algorithm}



\dotparagraph{Deposit procedure}
\begin{algorithm}[]
\caption{Client protocol for depositing on an EVM chain}
\label{alg:evm-chains}
\begin{algorithmic}[1]
    \State{var \var{bridgeValidators}} \Comment{The validators of \btm}
    \Statex

    \Function{\op{deposit(\var{token}, \var{amount}, \var{receiver}, \var{network}\varIsWrapped)}}{}
        \State{\op{depositErc20}(\var{token}, \var{amount}, \var{receiver}, \var{network}\varIsWrapped)}
        \State{\textbf{wait for} \event{DepositedErc20} event}
        \Let{\var{txHash}}{\textit{hash of the tx that contains the \op{depositErc20()} call}} \label{line:evm-identifier-begin}
        \Let{\var{txNonce}}{\textit{index of the \event{DepositedErc20} event in the logs of tx}}
        \Let{\var{chainID}}{\textit{identifier of source chain}}
        \Let{\var{depositIdentifier}}{\{\var{txHash}, \var{txNonce}, \var{chainID}\}}\label{line:evm-identifier-end}
        \For{\var{v} \textbf{in} \var{bridgeValidators}}
            \State{$\msgsubmitwithrawal{\var{depositIdentifier}} \rightarrow v$} \label{line:evm-call-tss} \Comment{see \Cref{alg:tss}}
        \EndFor
    \EndFunction{}
\end{algorithmic}
\end{algorithm}

The client initiates a bridging instance from an EVM chain by calling \op{depositErc20()} or \op{depositNative()} (for simplicity we only show the client code for ERC20 tokens, it is similar for native).
We present the logic in \Cref{alg:evm-chains}.
The protocol assumes that before executing the depositERC20 function the client has
(1) obtained information, such as available tokens, their addresses, and chain identifiers,
and (2) approved the contract to spend the desired amount of tokens.

When the corresponding event (\event{DepositedErc20} or \event{DepositedNative}) is generated, the client is responsible for constructing a \var{depositIdentifier}, as show in lines \ref{line:evm-identifier-begin}--\ref{line:evm-identifier-end}.
The client then informs the \btm validators about this deposit by calling \op{submitWithdrawal()}.

\subsection{Deposit on Bitcoin} \label{sec:deposit-bitcoin}
The client creates a bitcoin transaction with the following inputs and outputs:
\begin{itemize}
    \item An input with index $0$, which identifies the \emph{sender} of the bridging request (if seen as an output of some other transaction \var{prevTx}, the sender is the public key that locks that output in \var{prevTX}).
    \item An output with index \var{txNonce}, the value of which indicate the \emph{amount} of the bridging request. This output must be locked with the bitcoin address controlled by the \btm validators.
    \item An output with index $\var{txNonce} + 1$, containing the \emph{target chain} in an \var{OP\_RETURN} script operator.
\end{itemize}
Let \var{txHash} denote the hash of this transaction. The client submits the transaction to bitcoin and constructs
\var{depositIdentifier}, containing \var{txHash}, \var{txNonce}, and an identifier for the bitcoin main-net as \var{chainID}.

\subsection{Deposit on Zano}
In order to initiate a bridging request with Zano as the source chain, a client must submit a transaction of type \const{burnOperation}. The amount and identifier of burnt asset will be used as \var{amount} and \var{tokenAddr}, respectively. Moreover, the client must use the \var{serviceEntries} field of the transaction to encode the \var{targetAddr} and \var{targetChain} of the bridging request.

After the transaction becomes finalized, the client constructs \var{depositIdentifier} setting as \var{txHash} its hash, as \var{txNonce} the index in \var{serviceEntries} where the corresponding information has been written, and as \var{chainID} the identifier of the Zano chain.
\section{Withdrawal Implementation} \label{sec:withdrawal}
The final step to complete a bridging request is the submission of the withdrawal transaction.
On non-EVM chains, the validators are responsible for finalizing the request to the target chain,
so the clients wait until the corresponding transaction is published.
This section describes how withdrawal to EVM is performed.

Withdrawals to an EVM chain are enabled by a smart contract, deployed by \btm on the target chain.
The smart contract is called by validators at finalization phase (as we saw in \Cref{sec:finalization-phase}), and they can also be called directly by the client.

\subsection{Smart contracts}
In order to support EVM chains as target, a number of smart contracts are deployed to that chain, shown in \Cref{alg:evm-withdraw-contracts}. These allow clients to withdraw native source-chain tokens or arbitrary ERC20 tokens.
The contracts compute the \var{signHash} (lines \ref{line:evm-contract-erc20-sign-hash} and \ref{line:evm-contract-native-sign-hash}) in the same way the \btm validator computed it earlier, require that the given amount is positive and the token and receiver addresses are not empty, and use the \op{transfer()} operation of \type{erc20} (\cref{line:evm-contract-erc20-transfer}) or a native EVM call (\cref{line:evm-contract-native-transfer}) to transfer the amount to the recipient.

\begin{algorithm}[ht!]
\caption{Smart contracts for EVM target chains}
\label{alg:evm-withdraw-contracts}
\begin{algorithmic}[1]
    \Function{\op{withdrawErc20(}}{}
    \State{   \var{token}}, \Comment{address of token to be transferred}
    \State{   \var{amount}}, \Comment{amount of tokens to transfer}
    \State{    \var{receiver}}, \Comment{receiver address on the target network}
    \State{    \var{txHash}}, \Comment{hash of the tx that contains the deposit tx}
    \State{    \var{txNonce}}, \Comment{index of the event in the deposit tx}
    \State{   \var{signatures}},
    \State{)}       \Comment{\var{block}: the current executed block}
    \Let{\var{signHash}}{keccak256(\var{token}, \var{amount},  \var{receiver}, \var{txHash},  \var{txNonce}, \var{block}.\var{chainID})} \label{line:evm-contract-erc20-sign-hash}
    \Guard(\op{verifySignatures(\var{signatures}, \var{signHash})})  \Comment{\op{verifySignatures()} omitted}
    \Guard{\var{amount} > 0}
    \Guard{\var{token} \neq \bot}
    \Guard{\var{receiver} \neq \bot}
    \State{\type{erc20}.\op{transfer}(\var{receiver}, \var{amount})} \Comment{\type{erc20}.\op{transfer()} omitted}\label{line:evm-contract-erc20-transfer}
    \EndFunction
    \Statex

    \Function{\op{withdrawNative(}}{}
    \State{   \var{amount}}, \Comment{amount of tokens to transfer}
    \State{    \var{receiver}}, \Comment{receiver address on the target network}
    \State{    \var{txHash}}, \Comment{hash of the tx that contains the deposit tx}
    \State{    \var{txNonce}}, \Comment{index of the event in the deposit tx}
    \State{   \var{signatures}},
    \State{)}
    \Let{\var{signHash}}{keccak256(\var{amount},  \var{receiver}, \var{txHash},  \var{txNonce}, \var{block}.\var{chainID})} \label{line:evm-contract-native-sign-hash}
    \Guard(\op{verifySignatures(\var{signatures}, \var{signHash})})
    \Guard{\var{amount} > 0}
    \Guard{\var{receiver} \neq \bot}
    \State{\var{receiver}.\op{payable\_call}({\var{amount})}}\label{line:evm-contract-native-transfer}
    \EndFunction
\end{algorithmic}
\end{algorithm}


\subsection{Submitting the withdrawal}
The protocol of \btm allows client to directly submit a withdrawal transaction to the target chain, if the corresponding signature has been generated by the validators. We present the logic in \Cref{alg:evm-chains-withdraw}.
The client uses the \op{checkWithdrawal()} function exposed by the \btm validators (\cref{line:check-withdrawal} in \Cref{alg:tss}) to query the state of the bridging request identifier by \var{depositIdentifier} and to obtain the required data and the signature. The client then calls the contract in the target chain.
We observe here that the \op{withdraw()} function of \Cref{alg:evm-chains-withdraw} can be called by any client that has \var{depositIdentifier}, not necessarily by the client that initiated the bridging request.

\begin{algorithm}[ht!]
\caption{Client protocol for withdrawing on an EVM chain}
\label{alg:evm-chains-withdraw}
\begin{algorithmic}[1]
    \State{var \var{bridgeValidators}} \Comment{The validators of \btm}
    \Statex
    \Function{\op{withdraw(\var{depositId}})}{}
        \For{\var{validator} \textbf{in} \var{bridgeValidators}}
            \Let{(\var{depositData}, \var{sig})}{\var{validator}.\op{checkWithdrawal}(\var{depositId})}\label{line:evm-call-check-withdrawal} \Comment{see Alg. \ref{alg:tss}}
            \If{$\var{withdrawalData} \neq \bot$}
                \State{\op{withdrawErc20}(%
                    \var{depositData}.\var{amount},
                    \var{depositData}.\var{targetAddr},
                    \State{\qquad\qquad\qquad \var{depositId}.\var{txHash},}
                    \var{depositId}.\var{txNonce},
                    \var{depositId}.\var{chainID},
                    \var{sig})}
                    \State{\textbf{return}} \Comment{Similar for native withdrawals}
            \EndIf
        \EndFor
    \EndFunction
\end{algorithmic}
\end{algorithm}

\section{Chain Clients Implementation}\label{sec:chain-clients}
The \ChainClient module is used by \btm validators as a \emph{verifier} for events in the source chain, but also to create and submit transactions to the target chain.
A \ChainClient is designed for a specific chain. It exposes the following functions:
\begin{description}
    \item \op{getDepositData()}: It gets a \var{depositIdentifier}, verifies it on the corresponding chain, and, if successful, returns a variable of type \type{DepositData}.
    \item \op{getWithdrawalTx()}: It gets a \var{depositId} and the corresponding \type{DepositData} and returns an unsigned transaction for the target chain.
    \item \op{getHashOfWithdrawal()}: It returns the hash that needs to be signed for the withdrawal transaction.
    \item \op{submitTx()}: It gets a signed transaction, or a transaction and a signature, and submits it to the target chain.
\end{description}

\subsection{The chain client for EVM}
The EVM \ChainClient connects to an Ethereum RPC provider (the implementation uses Infura) to query the state of the blockchain. In the pseudocode (\Cref{alg:evm-chain-client}) this is represented with the \var{Rpc} variable.

\begin{algorithm}[ht!]
\caption{Validator code, chain-client for EVM}
\label{alg:evm-chain-client}
\begin{algorithmic}[1]
    \State{const \const{requiredConfirmationsEvm}}
    \State{var \var{Rpc}} \Comment{An RPC provider for the supported chain}

    \Statex
    \Function{$\op{getDepositData}(\op{\var{depositIdentifier}}) \rightarrow \type{DepositData}$}{}
        \Let{\var{tx}}{\var{Rpc}.\op{eth\_getTransactionByHash}(\var{depositIdentifier.txHash})} \label{line:evm-get-transaction}
        \Let{\var{receipt}}{\var{Rpc}.\op{eth\_getTransactionReceipt}(\var{depositIdentifier.txHash})} \label{line:evm-get-transaction-receipt}
        \Let{\var{curHeight}}{\var{Rpc}.\op{eth\_blockNumber()}} \label{line:evm-block-count}
        \Guard{\var{tx} \neq \error \textbf{ and } \var{receipt} \neq \error \textbf{ and } \var{curHeight} \neq \error}
        \Guard{\var{curHeight} \geq \var{\var{receipt}.\var{blockNum}} + \const{requiredConfirmationsEvm} - 1 } \label{line:evm-check-confirmations}
        \Let{\var{eventBody}}{\var{receipt}.\var{logs}[\var{depositIdentifier.txNonce}]} \textbf{as} \const{DepositedErc20} \label{line:evm-find-log-begin}
        \If{$\var{eventBody} \neq \error$}
            \Let{\var{tokenAddr}}{\var{eventBody.token}}
        \Else
            \Let{\var{eventBody}}{\var{receipt}.\var{logs}[\var{depositIdentifier.txNonce}]} \textbf{as} \const{DepositedNative}
            \Let{\var{tokenAddr}}{\const{nativeTokenAddress}}
            \EndIf\label{line:evm-find-log-end}
        \Guard{\var{eventBody} \neq \error}

        \Let{\var{amount}}{\var{eventBody}.\var{amount}}
        \Let{\var{sourceAddr}}{\var{tx}.\var{from}}
        \Let{\var{targetAddr}}{\var{eventBody}.\var{receiver}}
        \Let{\var{targetChain}}{\var{eventBody}.\var{network}}
        \return{\type{DepositData}\{\var{tokenAddr},
                \var{amount},
                \var{sourceAddr},
                \var{targetAddr},
                \var{targetChain}\}} \label{line:evm-return-deposit-data}
    \EndFunction
    \Statex



    \Function{\op{getHashOfWithdrawal(\var{depositId}, \var{depositData})}}{}
    \If{\var{depositData}.\var{tokenAddr} == \const{native}}
        \Let{\var{data}}{(
            \var{depositData}.\var{amount},
            \var{depositData}.\var{targetAddr},
            \var{depositId}.\var{txHash},}
            \State{\qquad\quad \var{depositId}.\var{txNonce},
            \var{depositId}.\var{chainID}%
        )}
    \Else
        \Let{\var{data}}{(
            \var{depositData}.\var{tokenAddr},
            \var{depositData}.\var{amount},
            \var{depositData}.\var{targetAddr},
            \var{depositId}.\var{txHash},}
            \State{\qquad\quad\var{depositId}.\var{txNonce},
            \var{depositId}.\var{chainID}%
        )}
    \EndIf
        \State{\textbf{return} \op{keccak256}(\var{data})}
    \EndFunction

    \Statex
    \Function{\op{submitTx(\var{depositId}, \var{depositData}, \var{sig})}}{}
    \If{\var{depositData}.\var{tokenAddr} == \const{native}}
        \State{\var{Rpc}.\op{withdrawNative}( %
                    \var{depositData}.\var{amount},
                    \var{depositData}.\var{targetAddr}, \Comment{defined in Alg. \ref{alg:evm-withdraw-contracts}}
                    \State{\qquad\qquad\qquad \var{depositId}.\var{txHash},}
                    \var{depositId}.\var{txNonce},
                    \var{depositId}.\var{chainID},
                    \var{sig})}
    \Else
        \State{\var{Rpc}.\op{withdrawErc20}(%
                    \var{depositData}.\var{amount},
                    \var{depositData}.\var{targetAddr}, \Comment{defined in Alg. \ref{alg:evm-withdraw-contracts}}
                    \State{\qquad\qquad\qquad \var{depositId}.\var{txHash},}
                    \var{depositId}.\var{txNonce},
                    \var{depositId}.\var{chainID},
                    \var{sig})}
    \EndIf
    \EndFunction
\end{algorithmic}
\end{algorithm}


Function \op{getDepositData()} takes a \var{depositIdentifier} and retrieves the corresponding deposit from an EVM chain.
It uses the RPC provider to retrieve the transaction \var{tx} and its receipt (which represents the results of a transaction in an EVM chain and includes the emitted events) indicated by \var{txHash} (lines \ref{line:evm-get-transaction}--\ref{line:evm-get-transaction-receipt}).
If any of these calls returns an error (e.g., because the transaction was not found), it aborts.
The transaction must have the required number of confirmations in the source chain, otherwise the \ChainClient also aborts (\cref{line:evm-check-confirmations}).

The \ChainClient then tries to retrieve from the receipt the event related to the deposit, either \event{DepositedErc20} or \event{DepositedNative}, using the given \var{txNonce} (lines~\ref{line:evm-find-log-begin}--\ref{line:evm-find-log-end}).
Observe that if none of the two events can be parsed, then it aborts.
Also observe that, in the case of an \event{EventDepositedErc20} event, the address of the bridged token is extracted from the event.
The rest of the fields of the returned \type{DepositData} are extracted from the \var{eventBody} (its format is shown in \Cref{alg:evm-deposit-contracts}) and from the transaction \var{tx} (\cref{line:evm-return-deposit-data}).

Finally, \op{getHashOfWithdrawal()} returns the hash of the appropriate fields, as expected by the EVM withdrawal contracts (see \cref{line:evm-contract-erc20-sign-hash} of \Cref{alg:evm-withdraw-contracts}).
Once the signature is obtained, validators can use \op{submitTx()} to submit the transaction to the target chain.

\begin{remark}
    In the end, \emph{all validators trust the same RPC provider}. There is no verification on the results returned by the RPC provider. The RPC provider is trusted for the safety and liveness of the protocol.
\end{remark}

\subsection{The chain client for Zano}
\label{sec:zano-chain-client}

\begin{algorithm}[ht!]
\caption{Validator code, chain-client for Zano}
\label{alg:zano-chain-client}
\begin{algorithmic}[1]
    \State{const \const{requiredConfirmationsZano}}
    \State{var \var{ZanoNode}} \Comment{A Zano full node }
    \Statex
    \Function{$\op{getDepositData}(\op{\var{depositIdentifier}}) \rightarrow \type{DepositData}$}{}
        \Let{\var{tx}}{\var{ZanoNode}.\op{getTransaction}(\var{depositIdentifier.txHash})} \label{line:zano-get-transaction}
        \Let{\var{curHeight}}{\var{ZanoNode}.\op{currentHeight()}} \label{line:zano-block-count}
        \Guard{\var{tx} \neq \error \textbf{ and } \var{curHeight} \neq \error}
        \Guard{\var{curHeight} \geq \var{\var{tx}.\var{height}} + \const{requiredConfirmationsZano}} \label{line:zano-check-confirmations}
        \Let{\var{sourceAddr}}{\var{tx}.\var{remoteAddresses[0]}}\label{line:zano-extract-begin}
        \Guard{\var{tx}.\var{ado}.\var{operationType} == \const{burnOperation}} \label{line:zano-require-burn}
        \Guard{\var{tx}.\var{ado}.\var{optAssetId} \neq \bot \textbf{ and } \var{tx}.\var{ado}.\var{optAmount} \neq \bot}
        \Let{\var{tokenAddr}}{\var{tx}.\var{ado}.\var{optAssetId}}
        \Let{\var{amount}}{\var{tx}.\var{ado}.\var{optAmount}}
        \Guard{\var{tx}.\var{serviceEntries[\var{depositIdentifier}.\var{txNonce}]} \neq \bot}
        \Let{(\var{targetAddr},\var{targetChain})}{\var{tx}.\var{serviceEntries[\var{depositIdentifier}.\var{txNonce}]}} \label{line:zano-extract-end}
        \return{\type{DepositData}\{\var{tokenAddr}, \var{amount}, \var{sourceAddr}, \var{targetAddr},
\var{targetChain}\}} \label{line:zano-return-deposit-data}
        \EndFunction
    \Statex

    \Function{\op{getWithdrawalTx(\var{depositId}, \var{depositData})}}{}
        \Let{\var{tx}}{\text{create a \var{emitAsset} transaction with recipient \var{depositData.targetAddr}}}
        \State{\qquad\quad \text{and amount \var{depositData.amount}}}
        \State{\textbf{return} \var{tx}}
    \EndFunction
    \Statex

    \Function{\op{getHashOfWithdrawal(\var{depositId}, \var{depositData})}}{}
        \Let{\var{tx}}{\op{getWithdrawalTx}(\var{depositId}, \var{depositData})}
        \State{\textbf{return} \op{hash(\var{tx})}}
    \EndFunction

    \Statex
    \Function{\op{submitTx(\var{tx}, \var{sig})}}{}
        \State{\var{ZanoNode}.\op{submitTransaction}(\var{tx}, \var{sig})}
    \EndFunction
\end{algorithmic}
\end{algorithm}

The \ChainClient for Zano is shown in \Cref{alg:zano-chain-client}.
Each validator runs a Zano full node, and has access to \emph{two} different keypairs:
(1) Transactions on the Zano chain are encrypted. For transactions related to \btm, the same public key is used for encryption, and all validators have access to the corresponding private key to decrypt the deposit transaction. This detail is abstracted in the \var{ZanoNode}.\op{getTransaction()} call in the pseudocode.
(2) This should not be confused with another keypair used for submitting withdrawal transactions to the Zano chain (as part of the \emph{signing} protocol in \Cref{sec:bridge-withdrawal-generation}). This private key is unknown to anyone; validators only hold shares of it.

In \op{getDepositData()}, the validator uses the Zano full node to retrieve the deposit transaction (\cref{line:zano-get-transaction}).
The deposit transaction must be a \emph{burn operation} on the Zano chain (\cref{line:zano-require-burn}).
The required fields are then extracted from the transaction (lines \ref{line:zano-extract-begin}--\ref{line:zano-extract-end}) to form a \type{DepositData} (\cref{line:zano-return-deposit-data}).

For bridging to Zano, the \op{getWithdrawalTx()} function creates a new \emph{emit operation} transaction, with the recipient address and amount specified in \type{DepositData}. Function \op{getHashOfWithdrawal()} returns the hash of this transaction. Function \op{submitTx()} submits the transaction to the Zano network.

\subsection{The chain client for Bitcoin}

\dotparagraph{Notation}
\sloppy{
Regarding notation, a bitcoin transaction \var{tx} contains its \emph{inputs}, also called \emph{previous outputs}, in a field \var{tx.vin}, and its \emph{outputs} in a field \var{tx.vout}.
Each input and output is a \emph{UTXO}. The field \var{in.value} or \var{out.value} contains the amount of the UTXO, and the field \var{in.scriptPubKey} or \var{out.scriptPubKey} contains a script that locks the UTXO. Notation \var{out.scriptPubKey.address} indicates the address of that script, and \var{out.scriptPubKey.OP\_RETURN} returns the data written in an \var{OP\_RETURN} operator, or \error if no such operator exists.
Moreover, the field \var{tx.confirmations} returned by a bitcoin full node indicates the number of confirmations for \var{tx}. If it is $0$ then \var{tx} is in the mempool, a positive value $1+$ indicates how many blocks deep the transaction is, $1$ being the most recent block, and a negative value indicates that the transaction is in a reorg block.
}

\begin{algorithm}[ht!]
\caption{Validator code, \ChainClient for bitcoin}
\label{alg:btc-chain-client}
\begin{algorithmic}[1]
    \State{bitcoin full node \var{FullNode}}
    \State{const int \const{requiredConfirmationsBtc}}
    \State{const \const{bridgeAddress}}

    \Statex
    \Function{$\op{getDepositData}(\op{\var{\var{depositId}}}) \rightarrow \type{DepositData}$}{}
        \Let{\var{tx}}{\var{FullNode}.\op{getTransaction(\var{\var{depositId}.txHash})}} \label{line:btc-get-transaction}
        \Guard{\var{tx} \neq \error \textbf{ and } \var{tx}.\var{confirmations} \geq \const{requiredConfirmationsBtc}}
        \Guard{\op{len}(\var{tx.vout}) \geq \var{\var{depositId}}.\var{txNonce} + 1}
        \Guard{\var{tx.vout[\var{\var{depositId}}.\var{txNonce}].scriptPubKey.address} == \const{bridgeAddress}}\label{line:btc-check-address}
        \Let{\var{amount}}{\var{tx.vout[\var{depositId}.\var{txNonce}].\var{value}}} \label{line:btc-extract-amount}
        \Let{(\var{targetChain}, \var{targetAddr})}{\var{tx.vout}[\var{\var{depositId}}.\var{txNonce} + 1].\var{OP\_RETURN}} \label{line:btc-extract-address}
        \Guard{\var{amount} > 0 \textbf{ and } \var{targetChain} \neq \error \textbf{ and } \var{targetAddr} \neq \error} \label{line:btc-check-error-address}
        \Let{\var{prevTxID}}{\var{tx.vin[0].txid}}
        \Let{\var{prevTxOut}}{\var{tx.vin[0].vout}}
        \Let{\var{prevTx}}{\var{FullNode}.\op{getTransaction}(\var{prevTxID})}
        \Let{\var{sourceAddr}}{\var{prevTx.vout[\var{prevTxOut}].scriptPubKey.address }}
        \Guard{\var{sourceAddr} \neq \error}
        \Let{\var{tokenAddr}}{\const{bitcoin}}
        \return{\type{DepositData}\{\var{tokenAddr},
                \var{amount},
                \var{sourceAddr},
                \var{targetAddr},
                \var{targetChain}\}} \label{line:btc-return-deposit-data}
    \EndFunction
    \Statex

    \Function{\op{getWithdrawalTx(\var{depositId}, \var{depositData})}}{}
        \Let{\var{tx}}{\text{new bitcoin transaction}}
        \Let{\var{receiverOut}}{\text{new UTXO }}
        \Let{\var{receiverOut}.\var{value}}{\var{depositData}.\var{amount}}
        \Let{\var{receiverOut}.\var{scriptPubKey}}{\op{payToAddr}(\var{depositData}.\var{targetAddr})} \Comment{\op{payToAddr}() omitted}
        \State{\var{tx}.\var{vout}.\op{append}(\var{receiverOut})}
        \Let{\var{tx}.\var{vin}}{\text{collect UTXOs for \const{bridgeAddress} excl. \var{lockedBitcoinInputs}}} \label{line:btc-collect-inputs} \Comment{details omitted}
        \State{\textbf{return} \var{tx}}
    \EndFunction
    \Statex

    \Function{\op{getHashOfWithdrawal(\var{depositId}, \var{depositData})}}{}
        \Let{\var{tx}}{\op{getWithdrawalTx}(\var{depositId}, \var{depositData})}
        \Let{\var{sighash}}{\emptyset}
        \For{\var{input} \textbf{in} \var{tx}.\var{vin}}
            \State{\var{sighash}.\op{append}(\op{computeSighash}(\var{input}))} \Comment{\op{computeSighash}() omitted}
        \EndFor
        \State{\textbf{return} \var{sighash}}
    \EndFunction

    \Statex
    \Function{\op{submitTx(\var{signedTx})}}{}
        \State{\var{FullNode}.\op{submitTransaction}(\var{signedTx})} \Comment{\op{submitTransaction}() omitted}
    \EndFunction

\end{algorithmic}
\end{algorithm}

We show the \ChainClient for bitcoin in \Cref{alg:btc-chain-client}.
The deposit verification (function \op{getDepositData()}) works as follows.
The node first uses its local bitcoin full node to obtain the transaction \var{tx} indicated by hash \var{depositIdentifier.txNonce}. If the transaction is not found, or it is not deep enough in the bitcoin chain, then the deposit request fails.
The code then checks whether the transaction was correctly formed by the client (as shown in \Cref{sec:deposit-bitcoin}) and extracts the required fields.
The deposited amount is taken from the output with index \var{txNonce} (\cref{line:btc-extract-amount}).
It is required that the public key, with which this output is locked, corresponds to the address that is controlled by the validators (\cref{line:btc-check-address}).
The target chain and target address are extracted from \var{OP\_RETURN} in the script of the output with index $\var{depositIdentifier}.\var{txNonce} + 1$, and the deposit fails if any error occurs (e.g., the output does not exist, or the script of the output does not contain an \var{OP\_RETURN} operator; or the contents of \var{OP\_RETURN} do not have the expected format) (lines~\ref{line:btc-extract-address}--\ref{line:btc-check-error-address}).
The function returns a \type{depositData} variable.

The code for generating a withdrawal transaction for bridging to bitcoin is shown in function \op{getWithdrawalTx()}. It creates a new bitcoin transaction with one output UTXO, whose value is the bridged amount. The script that locks this output is constructed from \var{targetAddr} (the implementation of \op{payToAddr()} is omitted).
The inputs to this transaction are collected from the unspent outputs available for the bridge address (\cref{line:btc-collect-inputs}), excluding UTXOs in \var{lockedBitcoinInputs}: it can be the case that some UTXOs appear as unspent in the mempool because an older withdrawal transaction has not yet reached the bitcoin network. The \var{lockedBitcoinInputs} variable gets updated at the end of the \emph{finalization} phase of the protocol (see \Cref{alg:finalization-phase}) with the inputs to each withdrawal transaction. We leave the exact logic for this as an implementation detail (observe this might result in more than one input UTXOs and in a change output UTXO).

Finally, the \op{getHashOfWithdrawal()} function, which is used by the validators to create the message to be signed in the signing phase, first gets the withdrawal transaction and then returns the \var{sighash} for each input.
This uses a function \op{computeSighash()}, which follows the bitcoin consensus rules to compute the message that needs to be signed in order to spend an input (implementation omitted).

\begin{remark} [Collecting input UTXOs to fund the withdrawal transaction]
    Given $\var{depositData}.\var{amount}$ and the $\var{depositData}.\var{targetAddr}$, the \op{getWithdrawalTx()} function must be deterministic, i.e., return the same transaction to each validator.
\end{remark}
\section{Conclusion}\label{sec:conclusion}

This paper describes \btm, a bridging protocol for transferring tokens across
different blockchains. We offer a precise pseudocode description of \btm and
show that it guarantees the needed bridge properties, namely safety and
liveness. Finally, we outline various implementation details of completing 
a bridging request and retrieving or publishing the necessary objects from/to
the supported blockchains. 

\paragraph{Future work.}

A major point of consideration for further research is introducing
accountability in the protocol. The main idea is to identify misbehaving
parties, when something goes wrong and the state machine does not progress as
expected, and then exclude these parties from the execution. For example, if a
proposer is corrupted and fails to broadcast a valid deposit's id or construct
the signing committee, or if a validator's actions result in a TSS error, then
they could be excluded from future sessions. By constructing a robust
accountability mechanism, which always identifies a misbehaving party but never
penalizes correct participants, liveness could be significantly improved by
decreasing the number of sessions needed to have a good success probability
(cf. \cref{fig:liveness-simulation}).

\iflncs
\thispagestyle{plain}
\fi

\ifccs
  \bibliographystyle{ACM-Reference-Format}
\else
  \bibliographystyle{alpha}
\fi

\pdfbookmark[section]{References}{references}
\bibliography{dblpbibtex,biblio,crypto,pubs}

\pagebreak
\newpage
\section*{About Common Prefix}
Common Prefix is a blockchain research, development, and consulting
company consisting of scientists and engineers
specializing in all aspects of blockchain science.

We work using rigorous cryptographic techniques to design and implement simple, provably secure protocols from first principles. Our consulting and auditing pertain to theoretical cryptographic protocol analyses as well as the pragmatic auditing of implementations in both core consensus technologies and application layer smart contracts.

\begin{figure}
\center{\includesvg[width=0.4\columnwidth]{figures/CP_LOGO.svg}}
\label{fig:cp_logo}
\end{figure}


\end{document}